\documentclass[aps,prc,twocolumn,amsfonts,superscriptaddress,showkeys,nofootinbib,floatfix]{revtex4-1}

\usepackage{url}
\usepackage{color}
\usepackage{amssymb}
\usepackage{amsmath}
\usepackage{graphics,epsfig}
\usepackage{verbatim}
\usepackage{amstext}
\usepackage{mathrsfs}
\usepackage{latexsym}
\usepackage[caption=false]{subfig}
\usepackage{wasysym}
\usepackage{fancyhdr}
\usepackage{epstopdf}
\usepackage{bm}
\usepackage[colorlinks,breaklinks]{hyperref}
\usepackage{gensymb}

\newcommand {\nd} {$^{15}$ND$_{3}$}
\newcommand {\nh} {$^{15}$NH$_{3}$}
\newcommand {\deep} {$^2\vec{{\mathrm H}}(\vec{e},e^{\prime}p)n$}

\newcommand {\bt} {$P_{b} P_{t}$}

\begin{document}

\title{Beam-target double spin asymmetry in quasi-elastic electron scattering off the deuteron with CLAS}
 
%%%%%%%%%%%% Institutes Number and defintions %%%%%%%%%%%

% FIRST time through establishes the order
% RUN AUTHOR LIST BEFORE GOING TO PWG REVIEW

%%%%%%%%%%%%%%%%%%%%%%%%%%%%%%%%%%% 
\newcommand*{\ODU}{ Old Dominion University, Norfolk, Virginia 23529}
\newcommand*{\ODUindex}{28}
\affiliation{\ODU}
\newcommand*{\NOWPNNL}{Pacific Northwest National Laboratory, Richland, WA 99354.}
\newcommand*{\ANL}{Argonne National Laboratory, Argonne, Illinois 60439}
\newcommand*{\ANLindex}{1}
\affiliation{\ANL}
\newcommand*{\ASU}{Arizona State University, Tempe, Arizona 85287-1504}
\newcommand*{\ASUindex}{2}
\affiliation{\ASU}
\newcommand*{\CSUDH}{California State University, Dominguez Hills, Carson, CA 90747}
\newcommand*{\CSUDHindex}{3}
\affiliation{\CSUDH}
\newcommand*{\CANISIUS}{Canisius College, Buffalo, NY}
\newcommand*{\CANISIUSindex}{4}
\affiliation{\CANISIUS}
\newcommand*{\CMU}{Carnegie Mellon University, Pittsburgh, Pennsylvania 15213}
\newcommand*{\CMUindex}{5}
\affiliation{\CMU}
\newcommand*{\CUA}{Catholic University of America, Washington, D.C. 20064}
\newcommand*{\CUAindex}{6}
\affiliation{\CUA}
\newcommand*{\SACLAY}{Irfu/SPhN, CEA, Universit\'e Paris-Saclay, 91191 Gif-sur-Yvette, France}
\newcommand*{\SACLAYindex}{7}
\affiliation{\SACLAY}
\newcommand*{\CNU}{Christopher Newport University, Newport News, Virginia 23606}
\newcommand*{\CNUindex}{8}
\affiliation{\CNU}
\newcommand*{\UCONN}{University of Connecticut, Storrs, Connecticut 06269}
\newcommand*{\UCONNindex}{9}
\affiliation{\UCONN}
\newcommand*{\UGEN}{Universit\`a di Genova, Dipartimento di Fisica, 16146 Genova, Italy}
\newcommand*{\UGENindex}{10}
\affiliation{\UGEN}
\newcommand*{\FU}{Fairfield University, Fairfield CT 06824}
\newcommand*{\FUindex}{11}
\affiliation{\FU}
\newcommand*{\FIU}{Florida International University, Miami, Florida 33199}
\newcommand*{\FIUindex}{12}
\affiliation{\FIU}
\newcommand*{\FSU}{Florida State University, Tallahassee, Florida 32306}
\newcommand*{\FSUindex}{13}
\affiliation{\FSU}
\newcommand*{\GWUI}{The George Washington University, Washington, DC 20052}
\newcommand*{\GWUIindex}{14}
\affiliation{\GWUI}
\newcommand*{\ISU}{Idaho State University, Pocatello, Idaho 83209}
\newcommand*{\ISUindex}{15}
\affiliation{\ISU}
\newcommand*{\INFNFE}{INFN, Sezione di Ferrara, 44100 Ferrara, Italy}
\newcommand*{\INFNFEindex}{16}
\affiliation{\INFNFE}
\newcommand*{\INFNFR}{INFN, Laboratori Nazionali di Frascati, 00044 Frascati, Italy}
\newcommand*{\INFNFRindex}{17}
\affiliation{\INFNFR}
\newcommand*{\INFNGE}{INFN, Sezione di Genova, 16146 Genova, Italy}
\newcommand*{\INFNGEindex}{18}
\affiliation{\INFNGE}
\newcommand*{\INFNRO}{INFN, Sezione di Roma Tor Vergata, 00133 Rome, Italy}
\newcommand*{\INFNROindex}{19}
\affiliation{\INFNRO}
\newcommand*{\INFNTUR}{INFN, Sezione di Torino, 10125 Torino, Italy}
\newcommand*{\INFNTURindex}{20}
\affiliation{\INFNTUR}
\newcommand*{\ORSAY}{Institut de Physique Nucl\'eaire, CNRS/IN2P3 and Universit\'e Paris Sud, Orsay, France}
\newcommand*{\ORSAYindex}{21}
\affiliation{\ORSAY}
\newcommand*{\ITEP}{Institute of Theoretical and Experimental Physics, Moscow, 117259, Russia}
\newcommand*{\ITEPindex}{22}
\affiliation{\ITEP}
\newcommand*{\JMU}{James Madison University, Harrisonburg, Virginia 22807}
\newcommand*{\JMUindex}{23}
\affiliation{\JMU}
\newcommand*{\KNU}{Kyungpook National University, Daegu 702-701, Republic of Korea}
\newcommand*{\KNUindex}{24}
\affiliation{\KNU}
\newcommand*{\MISS}{Mississippi State University, Mississippi State, MS 39762-5167}
\newcommand*{\MISSindex}{25}
\affiliation{\MISS}
\newcommand*{\UNH}{University of New Hampshire, Durham, New Hampshire 03824-3568}
\newcommand*{\UNHindex}{26}
\affiliation{\UNH}
\newcommand*{\NSU}{Norfolk State University, Norfolk, Virginia 23504}
\newcommand*{\NSUindex}{27}
\affiliation{\NSU}
\newcommand*{\OHIOU}{Ohio University, Athens, Ohio  45701}
\newcommand*{\OHIOUindex}{28}
\affiliation{\OHIOU}
%\newcommand*{\ODU}{Old Dominion University, Norfolk, Virginia 23529}
%\newcommand*{\ODUindex}{28}
%\affiliation{\ODU}
\newcommand*{\RPI}{Rensselaer Polytechnic Institute, Troy, New York 12180-3590}
\newcommand*{\RPIindex}{29}
\affiliation{\RPI}
\newcommand*{\URICH}{University of Richmond, Richmond, Virginia 23173}
\newcommand*{\URICHindex}{30}
\affiliation{\URICH}
\newcommand*{\ROMAII}{Universita' di Roma Tor Vergata, 00133 Rome Italy}
\newcommand*{\ROMAIIindex}{31}
\affiliation{\ROMAII}
\newcommand*{\MSU}{Skobeltsyn Institute of Nuclear Physics, Lomonosov Moscow State University, 119234 Moscow, Russia}
\newcommand*{\MSUindex}{32}
\affiliation{\MSU}
\newcommand*{\SCAROLINA}{University of South Carolina, Columbia, South Carolina 29208}
\newcommand*{\SCAROLINAindex}{33}
\affiliation{\SCAROLINA}
\newcommand*{\TEMPLE}{Temple University,  Philadelphia, PA 19122 }
\newcommand*{\TEMPLEindex}{34}
\affiliation{\TEMPLE}
\newcommand*{\JLAB}{Thomas Jefferson National Accelerator Facility, Newport News, Virginia 23606}
\newcommand*{\JLABindex}{35}
\affiliation{\JLAB}
\newcommand*{\UTFSM}{Universidad T\'{e}cnica Federico Santa Mar\'{i}a, Casilla 110-V Valpara\'{i}so, Chile}
\newcommand*{\UTFSMindex}{36}
\affiliation{\UTFSM}
\newcommand*{\EDINBURGH}{Edinburgh University, Edinburgh EH9 3JZ, United Kingdom}
\newcommand*{\EDINBURGHindex}{37}
\affiliation{\EDINBURGH}
\newcommand*{\GLASGOW}{University of Glasgow, Glasgow G12 8QQ, United Kingdom}
\newcommand*{\GLASGOWindex}{38}
\affiliation{\GLASGOW}
\newcommand*{\VIRGINIA}{University of Virginia, Charlottesville, Virginia 22901}
\newcommand*{\VIRGINIAindex}{39}
\affiliation{\VIRGINIA}
\newcommand*{\WM}{College of William and Mary, Williamsburg, Virginia 23187-8795}
\newcommand*{\WMindex}{40}
\affiliation{\WM}
\newcommand*{\YEREVAN}{Yerevan Physics Institute, 375036 Yerevan, Armenia}
\newcommand*{\YEREVANindex}{41}
\affiliation{\YEREVAN}

\newcommand*{\NOWJLAB}{Thomas Jefferson National Accelerator Facility, Newport News, Virginia 23606}
%\altaffiliation{\NOWPNNL}
\newcommand*{\NOWVCU}{Virginia Commonwealth University, Richmond, Virginia 23284}

%%%%%%%%%%%%%%%%%%%% authors %%%%%%%%% 
%\preprint{WM-06-xxx}
\author{M.~Mayer}
\altaffiliation{\NOWPNNL}
\affiliation{\ODU}
\author {S.E.~Kuhn} 
     \thanks{Corresponding author. Email: skuhn@odu.edu}
\affiliation{\ODU}
%\author{P.~Bosted}
%\affiliation{\WM}
%\author{C.~Keith}
%\affiliation{\JLab}
%\author{D.~Meekins}
%\affiliation{\JLab}
\author {Z.~Akbar} 
\affiliation{\FSU}
\author {S. ~Anefalos~Pereira} 
\affiliation{\INFNFR}
\author {G.~Asryan} 
\affiliation{\YEREVAN}
\author {H.~Avakian} 
\affiliation{\JLAB}
\affiliation{\INFNFR}
\author {R.A.~Badui} 
\affiliation{\FIU}
\author {J.~Ball} 
\affiliation{\SACLAY}
\author {N.A.~Baltzell} 
\affiliation{\JLAB}
\author {M.~Battaglieri} 
\affiliation{\INFNGE}
% ...

\collaboration{The CLAS Collaboration}
\noaffiliation

 \date{\today}

\begin{abstract}
\begin{description}
\item[Background]
The deuteron plays a pivotal role in nuclear and hadronic physics, as both the simplest bound multi-nucleon system and as an ``effective neutron target''. Quasi-elastic electron scattering on the deuteron is a benchmark reaction to test our understanding of deuteron structure and the properties and interactions of the two nucleons bound in the deuteron.
\item[Purpose]
The experimental data presented here can be used to test state-of-the-art models of the deuteron and the two-nucleon interaction in the final state after two-body breakup of the deuteron. Focusing on polarization degrees of freedom, we gain information on spin-momentum correlations in the deuteron ground state (due to the D-state admixture) and on the limits of the Impulse Approximation (IA) picture as it applies to measurements of spin-dependent observables like spin structure functions for bound nucleons. Information on this reaction can also be used to reduce systematic uncertainties on the determination of neutron form factors or deuteron polarization through  quasi-elastic polarized electron scattering.
\item[Method]
We measured the beam-target double spin asymmetry ($A_{||}$) for quasi-elastic electron scattering off the deuteron at several beam energies (1.6$-$1.7 GeV, 2.5 GeV, 4.2 GeV and 5.6$-$5.8 GeV), using the CEBAF Large Acceptance Spectrometer (CLAS) at the Thomas Jefferson National Accelerator Facility.  The deuterons were polarized along (or opposite to) the beam direction.  The double spin asymmetries were measured as a function of photon virtuality $Q^{2}$ (0.13$-$3.17 (GeV$/c$)$^2$), missing momentum ($p_m = 0.0 - 0.5$ GeV$/c$), and the angle between the (inferred) ``spectator'' neutron and the momentum transfer direction ($\theta_{nq}$). 
\item[Results]
The results are compared with a recent model that includes Final State Interactions (FSI) using a complete parameterization of nucleon-nucleon scattering, as well as a simplified model using the Plane Wave Impulse Approximation (PWIA).  We find overall good agreement with both the PWIA and FSI expectations at low to medium missing momenta ($p_{m} \le 0.25$ GeV/$c$), including the change of the asymmetry due to the contribution of the deuteron D-state at higher momenta. At the highest missing momenta, our data clearly agree better with the calculations including FSI.
\item[Conclusions]
Final state interactions seem to play a lesser role for polarization observables in deuteron two-body electro-disintegration than for absolute cross sections. Our data, while limited in statistical power, indicate that PWIA models work reasonably well to understand the asymmetries at lower missing momenta. In turn, this information can be used to extract the product of beam and target polarization (\bt) from quasi-elastic electron-deuteron scattering, which is useful for measurements of spin observables in electron-neutron inelastic scattering. However, at the highest missing (neutron) momenta, FSI effects become important and must be accounted for.
\end{description}
\end{abstract}


\maketitle
\section{INTRODUCTION}\label{section:introduction}
The deuteron, as the simplest nuclear system, serves the dual role of an ``effective free neutron target''~\cite{Eden:1994ji, Abe:structure, Anthony:1999rm,  Alexakhin:2006oza, Airapetian:2006vy,Prok:2014ltt,Tkachenko:2014byy, Guler:2015hsw} and as a testing ground for sophisticated models of nucleon-nucleon interactions and scattering mechanisms~\cite{Arenhovel:2004bc,PhysRevC.45.2094}.  Electron scattering off the deuteron has been used as a means to extract information on its nuclear structure, including the $D$-wave ($L = 2$) contribution to the ground state wave function~\cite{ericson1983deuteron,forest1996femtometer}.  On the other hand, experiments that look for modifications of nucleon structure due to nuclear binding have also used the deuteron as a testbed~\cite{Klimenko06,Griffioen:2015hxa, Larry12}. In all of these cases, a thorough and detailed understanding of the scattering mechanism is necessary.

In particular, quasi-elastic scattering off the deuteron has been widely  studied~\cite{Ulmer:2002jn,hasell2011spin} as an ideal reaction to disentangle various contributions to the reaction mechanism, such as relativistic effects, non-nucleonic components of the deuteron wave function, meson-exchange (MEC) and isobar (IC) currents, and final state interactions (FSI) between the outgoing nucleons.  Recent experiments~\cite{Egiyan:2007qj,Boeglin:2011mt} have focused on higher momentum transfers, where one-nucleon currents are expected to dominate the cross section. Because of the continuing (and growing) importance of the deuteron as an effective neutron 
target~\cite{bonus12, eg12}, a particularly important question is whether there is a kinematic region where the simple picture of the  Plane Wave Impulse Approximation (PWIA) works reasonably well, in which the virtual photon is absorbed by only one nucleon inside the deuteron while the other is an unperturbed spectator to the reaction. Alternatively, one wants to test state of the art models of FSI to ascertain if they can yield a reliable description of the reaction mechanism.  In this quest, polarization degrees of freedom are particularly interesting, yet few experiments exist.  

From a practical point of view, quasi-elastic scattering off a polarized deuteron target (with or without detection of a final state proton) is often used as a direct measure of the product of beam and target polarization for spin structure function experiments~\cite{Dharmawardane:2006aa,Guler:2015hsw,Prok:2014ltt}. This requires that the theoretical asymmetry for this process is well-known, an assumption that should be tested experimentally.

In the following, we first give a brief overview of the theoretical background for the reaction $^2$H$(e,e^\prime p)n$, followed by an overview of existing data. Section~\ref{section:experiment} describes the experimental setup, followed by details of the data analysis and our results. The final section summarizes our findings.

%%%%%%%%%%%%%%%%%%%%%%%%%%%%%%%%%%%%%%%%%%%%%%%%%

\section{THEORETICAL BACKGROUND}\label{section:theory}
The deuteron is the simplest stable nucleus, consisting of a proton and neutron bound by only 2.2 MeV (see, for instance, the review by Gar\c con and Van Orden~\cite{Garcon:2001sz}). Its structure is amenable to detailed and sophisticated microscopic calculations that range from nonrelativistic approaches~\cite{forest1996femtometer}, based on the Schr\"odinger equation, to fully relativistic treatments~\cite{PhysRevC.45.2094}. Comparison of these calculations with experiment allows us to constrain properties of nucleons and of the nucleon-nucleon potential. In turn, given a model for the nucleon-nucleon interaction, form factors and momentum or spatial distributions of the nucleons in deuterium can be obtained~\cite{carlson:rmp98}.  Most modern models of the deuteron wave function agree in the basic features of this momentum distribution: At low momenta, it is dominated by the S-wave ($L = 0$) where the proton and neutron spin are parallel to the overall deuteron spin, while at momenta beyond 250-300 MeV/$c$, the contribution from the much smaller D-wave ($L=2$) component (which overall accounts for about 4-6\% of the deuteron ground state) becomes more important. In that kinematic region, the expectation value for the nucleon spins is actually opposite to that of the deuteron as a whole.

Experimentally, a large body of data on the quasi-elastic deuteron break-up reaction, $^2{\mathrm H}(e,e^{\prime}p)n$, has been collected to access information on the nucleon momentum distribution in deuterium (recent examples can be found in~\cite{passchier,hasell2011spin,Ulmer:2002jn,Egiyan:2007qj,Boeglin:2011mt,Boeglin:2014aca}). A very important question in this context is how the measured (missing) momentum distributions can be connected to the deuteron wave function~\cite{Ford:2014yua}, given their potential distortion by Final State Interactions (FSI)~\cite{Bianconi:1994bx}.

On the other hand, deuteron targets are often used to extract information on the neutron, due to the absence of sufficiently dense free neutron targets. For example, both unpolarized (see~\cite{Airapetian:2011nu} and references therein) and polarized~\cite{Abe:structure,Anthony:1999rm,Alexakhin:2006oza, Airapetian:2006vy,Dharmawardane:2006aa,Prok:2014ltt} structure functions of the neutron are often extracted from measurements on the deuteron.  In particular in the latter case, a clear understanding of the spin-dependent momentum distribution of nucleons in deuterium is of great importance, not only for a reliable extraction of neutron spin structure functions but also because the product of target and beam polarization (which enters the measured asymmetries as a constant factor) is often extracted using the polarized quasi-elastic reaction \deep~\cite{Dharmawardane:2006aa,Guler:2015hsw,Prok:2014ltt}.  Similarly, the $^2{\mathrm H}(e,e^{\prime}n)p$ reaction (with and without polarization information) is often used to access the neutron form factors~\cite{Zhu:2001md,Plaster:2005cx,Lachniet:2009aa}.  Furthermore, the novel technique of ``spectator tagging", where a backward-moving proton spectator is detected in coincidence with an inelastically scattered electron, is being used both to access the free neutron structure (at small spectator momenta~\cite{Baillie:2012ab,Tkachenko:2014aa,bonus12}) and to study possible modifications of nucleons that are part of a high-momentum correlation~\cite{Klimenko06,Larry12}. In all these cases, it is imperative to understand both the underlying spin-momentum structure of the deuteron as well as the reaction mechanism for electron scattering, including FSI effects.

The present paper focuses on the reaction \deep\ with a deuteron target polarized along the direction of the incoming electron beam. The differential cross section for this reaction,
\begin{equation}
\frac{d \sigma}{d  Q^2 d \phi_e d^3 \vec{p}_n} \equiv \sigma, 
\end{equation}
is a function of the (negative of the) squared four-momentum transferred by the scattered electron,
\begin{equation}
Q^2 = -(k-k^\prime)^2 = 4 E E^\prime \sin^2(\theta_e /2),
\end{equation}
with $k = (E, 0, 0, E)$ and 
\begin{equation}
k^\prime = (E^\prime, E^\prime \sin\theta_e \cos\phi_e,
E^\prime \sin\theta_e \sin\phi_e,E^\prime \cos\theta_e)
\end{equation}
being the four-momenta of the incoming and scattered electron (in the ultra relativistic limit), respectively. 
Here, $E$ is the energy of the incoming electron and $E^\prime$ is the energy of the scattered electron, while 
the scattered electron direction is given by  the polar angle $\theta_e$ and the azimuthal angle $\phi_e$
 with respect to the incoming electron beam.
The cross section also depends on the missing momentum $\vec{p}_m \equiv \vec{p}_n$ of the unobserved (but inferred) final state neutron; we will parametrize this momentum by its magnitude, $p_{m}$ and its angle $\theta_{nq}$ relative to the direction of the three-momentum transfer
$\vec{q} = \vec{k} - \vec{k}^\prime$.
For polarized beam and target, this cross section can be expressed as
\begin{equation}
 \sigma=\sigma_{0} \left[1+{\scriptstyle \sqrt{\frac{3}{2}}} P_{z} \left( A^{V}_{d} + h A^{V}_{ed} \right)
  +{\scriptstyle \sqrt{\frac{1}{2}}} P_{zz} \left( A^{T}_{d} + h A^{T}_{ed} \right) \right],
\label{poldeut:eq}
\end{equation}
where $\sigma_{0}$ is the unpolarized cross section, 
\begin{align}
P_z = \frac{N(+1) - N(-1)}{N(+1) + N(0) + N(-1)} \in \left[ -1, +1 \right] 
\end{align}
is the vector polarization and
\begin{align}
P_{zz} = \frac{N(+1) - 2 N(0) + N(-1)}{N(+1) + N(0) + N(-1)} \in \left[ -2, +1 \right] 
\end{align}
is the tensor polarization of the target (with $N(0,\pm1)$ the occupation numbers for the three magnetic quantum numbers $m_s = (+1, 0, -1)$), and $h$ is the helicity of the electrons.  We adopt here the notation of~\cite{Orden:Observables}, where the vector ($A^{V}_{d}$, $A^{V}_{ed}$) and tensor ($A^{T}_{d}$, $A^{T}_{ed}$) asymmetries are normalized as components of spherical tensors of rank 1 and 2, respectively.  Integration over all azimuthal directions $\phi_n$ of the final state neutron (around $\vec{q}$) leaves only the asymmetries $A^{V}_{ed}$ and $A^{T}_{d}$ (because of parity conservation in the electromagnetic interaction). Both asymmetries are functions of the beam energy $E$ and $Q^2$ as well as $p_{m}$ and $\cos\theta_{nq}$. Forming the difference between opposite-sign and equal sign pairs of helicity $h$ and target polarization $P_z$ and dividing by the sum, we arrive at the double-spin asymmetry
\begin{equation}
\label{eq:apara}
A_{||} = \frac{(\sigma_{-+} + \sigma_{+-}) - (\sigma_{--} + \sigma_{++})}
{\sigma_{-+} + \sigma_{+-}+\sigma_{--} + \sigma_{++}} = - 
\frac{\sqrt{3} P_{b} P_t A^{V}_{ed}}{\sqrt{2} +  P_{zz} A^{T}_{d}} ,
\end{equation}
where $P_b$ is the magnitude of the electron beam polarization and $P_t \equiv |P_z|$ is the average magnitude of the target vector polarization, both along the beam direction. The target used in the present experiment was vector-polarized up to $P_t \approx 0.4$ using Dynamical Nuclear Polarization (DNP)\cite{crabb:1997}, which yields a tensor polarization $P_{zz} \approx 0.1$, according to Equal Spin Temperature (EST) theory\cite{Abragam:1978}.
 
The simplest model for quasi-elastic deuteron breakup, the plane wave impulse approximation (PWIA), assumes that the virtual photon is absorbed by a single (on-shell) nucleon (impulse approximation) and the struck nucleon leaves the nucleus without further interaction (i.e., as a plane wave).  In this model, the measured asymmetry is proportional to the initial polarization of the struck
nucleon (ignoring the small contribution from the tensor asymmetry for the moment):
\begin{equation}
\label{eq:elasym}
A_{||} = \frac{P_{||} \sqrt{1 - \epsilon^2} + P_\perp \sqrt{2 \epsilon(1-\epsilon)} \frac{2 M}{\sqrt{Q^2}}
\frac{G_E}{G_M}}{1 + \epsilon \frac{4 M^2}{Q^2}\frac{G_E}{G_M}} .
\end{equation}
Here, $\frac{G_E}{G_M}$ is the ratio of electric to magnetic Sachs form factors of the struck nucleon, $P_{||}$ and $P_\perp$ are its polarization components along and transverse to the momentum transfer vector $\vec{q}$ (in the electron scattering plane), $M$ is the nucleon mass, and 
\begin{equation}
\label{epsilon:eq}
\epsilon  =  \left( 1 + 2[1 +
  \frac{Q^2}{4 M^2}]\tan^2\frac{\theta_e}{2}\right)^{-1} .
\end{equation}
is the virtual photon polarization ratio.  Within this PWIA picture, measurements of $A_{||}$ can  be used to extract information on the spin and momentum dependence of the nuclear wave function.  One goal of the present experiment is to determine the kinematic region where PWIA is a reasonably good approximation.

A more realistic description requires a model that includes the interaction between the spectator and knocked-out nucleon (FSI). Sabine Jeschonnek and J.W. Van Orden have developed a comprehensive theoretical model~\cite{Orden:Observables} for this purpose.  The authors model the \deep~reaction in various approximations (including PWIA) as well as with a complete treatment of FSI to learn more about its sensitivity to the initial nuclear state and the reaction mechanism.  They use a relativistic deuteron wave function by solving the Gross equation~\cite{PhysRevC.45.2094}.  A current SAID parameterization~\cite{SAIDdata} of the nucleon-nucleon scattering amplitude is used to calculate the interaction of the two nucleons in the final state, up to kinetic energies of about 1.3 GeV in the lab frame.  This nucleon-nucleon amplitude includes central, spin-orbit, and double spin-flip terms.  Within the kinematic range of applicability, we compare this model directly to our data, including the effects for both the vector and the tensor asymmetries in Eq.~(\ref{eq:apara}). 

%%%%%%%%%%%%%%%%%%%%%%%%%%%%%%%%%%%%%%%%%%%%%%%%%

\section{EXISTING DATA OVERVIEW}\label{section:wdata}
\begin{figure}[h!]
\begin{center}
\includegraphics[width=0.35\textwidth]{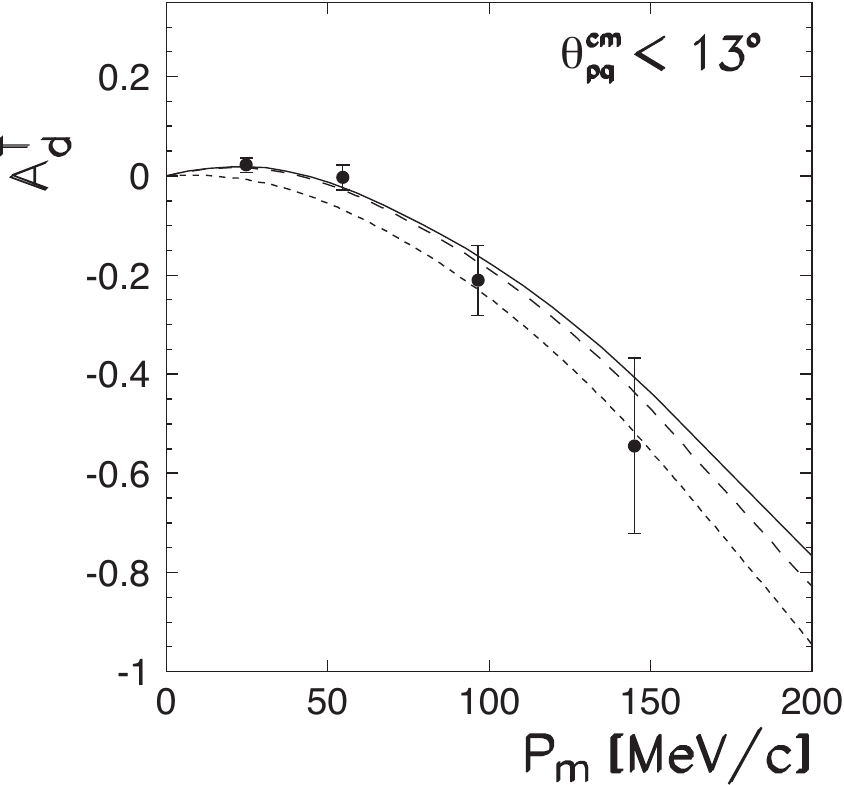}
\includegraphics[width=0.35\textwidth]{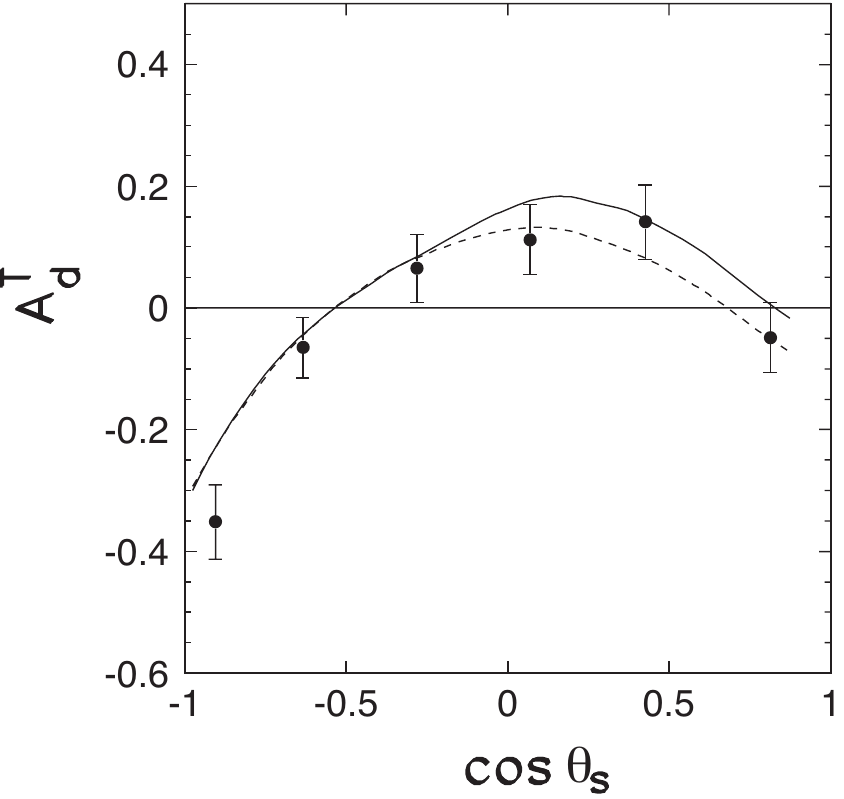}
  \end{center}
\caption[$A^T_{d}$ Results from NIKHEF]{Plots of $A^T_{d}$ from NIKHEF.  Theoretical curves from Arenh\"{o}vel are also shown.  Short-dashed curves are results for Plane Wave Born Approximation (PWBA), long-dashed curves include FSIs, and solid curves represent the full calculation~\cite{zhou1999}.}
\label{fig:nikehfat}
\end{figure}

Although the $^2$H$(e,e^{\prime}p)n$ reaction has been studied in detail, there exist only a few measurements of the beam-vector asymmetry $A^V_{ed}$ and tensor asymmetry $A^T_{d}$.  These asymmetries are directly related to the double spin asymmetry $A_{||}$ as seen in Eq.~(\ref{eq:apara}).  The existing data are at a relatively low $Q^2$ and were compared to a model formulated by Arenh\"{o}vel \textit{et al.}~\cite{Arenhovel:2004bc,arenhovel1995}.  This section will summarize the results of these experiments. In contrast, the new data reported in the following sections cover a much wider range in kinematics (beam energy and $Q^2$), and can therefore test models of FSI and deuteron structure in a region where different reaction mechanisms dominate.

\begin{figure}[!h]
\begin{center}
  \includegraphics[width=0.4\textwidth]{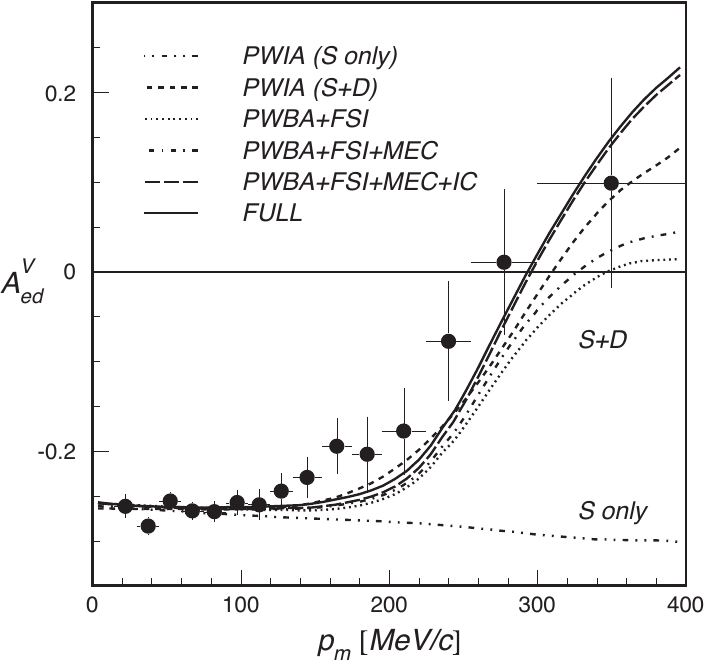}    
  \end{center}
  \caption[NIKHEF Results of $A_{ed}^{V}$]{NIKHEF results for $A_{ed}^{V}$ as a function of missing momentum at $Q^2$ = 0.21 (GeV/$c)^2$~\cite{passchier}.}
  \label{aved:fig}
\end{figure}

\subsection{NIKHEF}
\label{nikhef:wdata}
The first measurements of the tensor asymmetry $A^T_{d}$ were performed at the Dutch National Institute for Nuclear Physics and High Energy Physics (NIKHEF).  The experiment at NIKHEF used a polarized gas target with a 565 MeV electron beam~\cite{zhou1999}.  The tensor-polarized deuterium gas was altered between a polarization of $P^{+}_{zz}$=+0.488$\pm$0.014 and $P^{-}_{zz}$=-0.893$\pm$0.027 every 10 seconds.  Scattered electrons and protons were detected by the BigBite magnetic spectrometer.  The tensor asymmetry was extracted as a function of the angle $\theta_s$ between the polarization axis and the missing momentum, as well as a function of the magnitude of the missing momentum.  The range of missing momentum was limited to below 150 MeV/c.  The results of this measurement can be seen in Fig.~\ref{fig:nikehfat}.

Additionally, the first measurements of $A^V_{ed}$ were performed at NIKHEF several years later~\cite{passchier}.  A longitudinally polarized beam of electrons of 720 MeV was scattered off a vector-polarized deuterium target.  The scattered electron was measured at a fixed angle $\theta=40^{\circ}$, with a solid angle coverage of 96 millisteradians (msr) and knocked-out protons were measured at a central angle of $\theta_{p}=40^{\circ}$ with a solid angle coverage of 250 msr.  The missing momentum range was increased up to 350 MeV/$c$  at a $Q^2$ of 0.21 (GeV/$c)^2$.  Figure~\ref{aved:fig} shows that at momenta higher than 200 MeV/$c$, the vector asymmetry, $A_{ed}^{V}$, becomes sensitive to the $D$-state of the deuteron wave function.  

\begin{figure}
\begin{center}
 \includegraphics[width=0.48\textwidth]{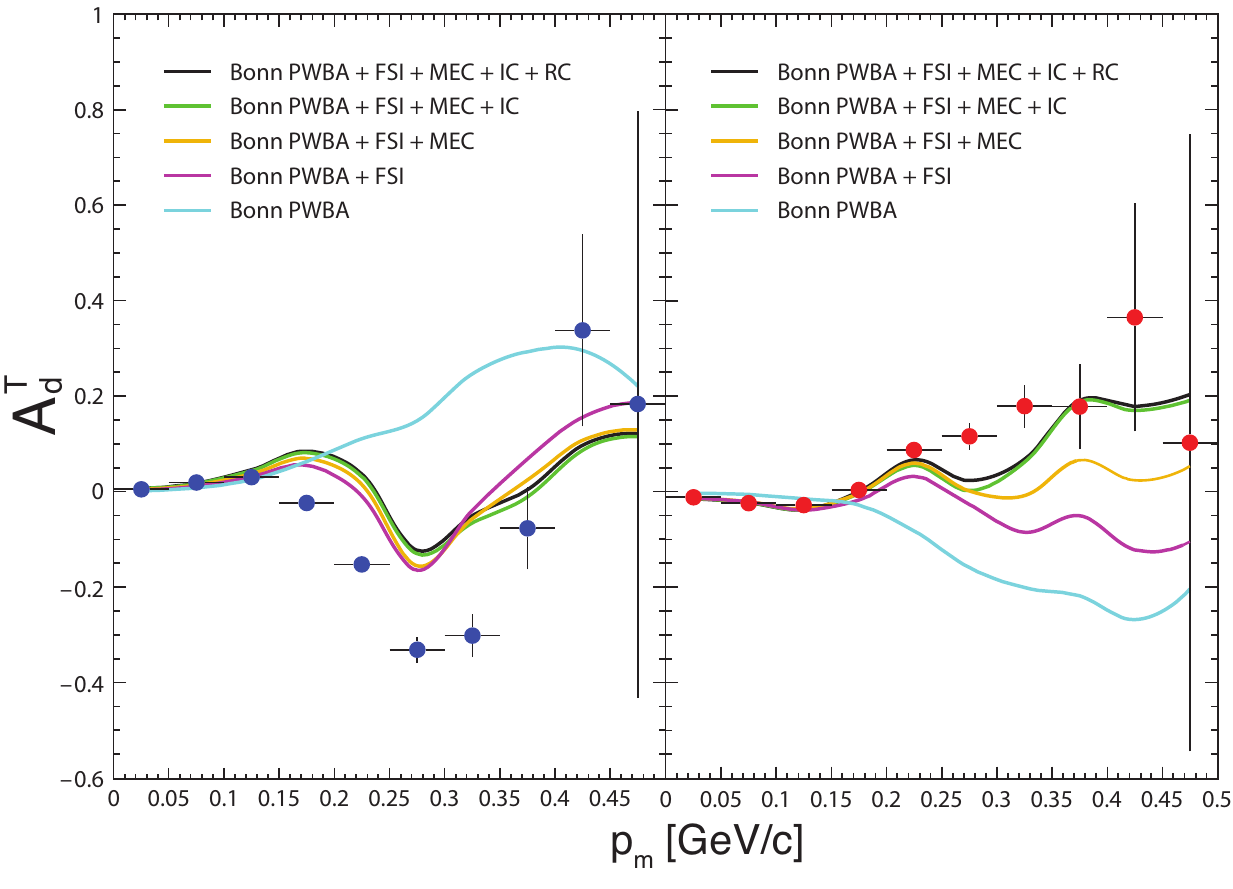}    
 \includegraphics[width=0.48\textwidth]{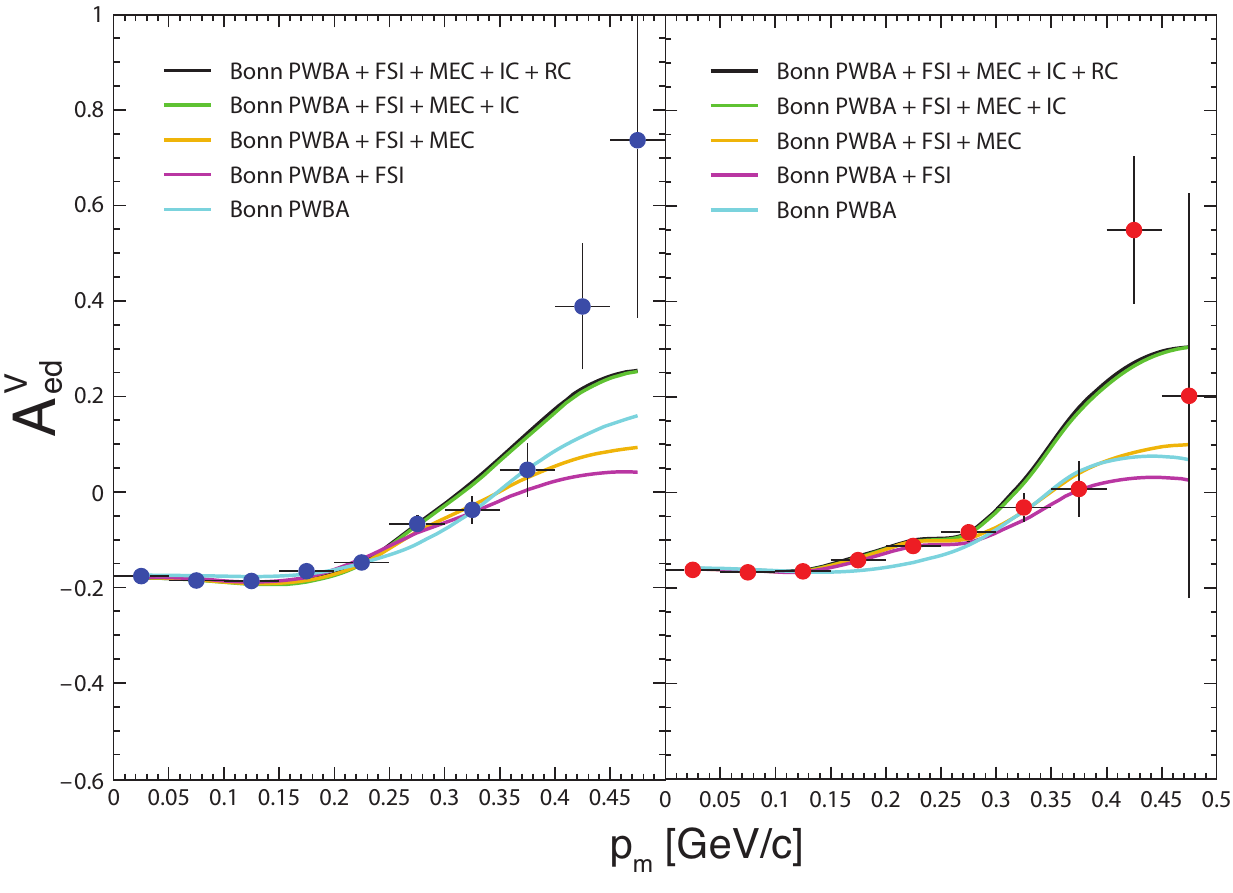}    
\end{center}
\caption[Results from BLAST]{Results from BLAST for A$^T_{d}$ (top) and A$^V_{ed}$ (bottom) for 0.2$<Q^2<$0.3 (GeV/c)$^2$, for both parallel (left) and perpendicular (right) kinematics.  Theoretical curves have been calculated including meson exchange currents (MEC), isobar currents (IC), and relativistic correction (RC)~\cite{degrush2010}.}
\label{fig:BLAST}
\end{figure}

\subsection{Bates}
The Bates Large Acceptance Spectrometer Toroid (BLAST) experiment used a polarized electron beam\footnote{Beam polarization $\sim60\%$ at 850 MeV.} incident upon an internal polarized deuterium target~\cite{Hasell:2011zz,hasell2011spin} at the MIT-Bates accelerator.  An atomic beam source was used for the polarized deuterium target, providing considerable freedom in the choice of vector and tensor polarization states.  Two sets of deuteron data were taken with nominal spin angles of 32$^\circ$ and 47$^\circ$ to provide perpendicular ($\theta^\ast,\phi^\ast = \pi/2,0$) and parallel ($\theta^\ast,\phi^\ast = 0,0$) kinematics\footnote{$(\theta^\ast, \phi^\ast$) describe the angle of the target polarization quantization axis relative to the momentum transfer vector, $\vec{q}$.}.  There are two analyses of the BLAST data; the latest work by A. DeGrush~\cite{degrush2010} re-evaluated the work of A. Maschinot~\cite{maschinot2005analysis} to extract $A^V_{ed}$ and $A^T_{d}$.  These data were taken for a $Q^2$ range of 0.1$<Q^2<$0.5 (GeV/c)$^2$ and ten missing momentum bins from 0.0 to 0.5 GeV/c.  The results of this measurement can be seen in Fig.~\ref{fig:BLAST}.

\begin{figure}[tbh]
\begin{center}    
  \includegraphics[width=0.32\textwidth]{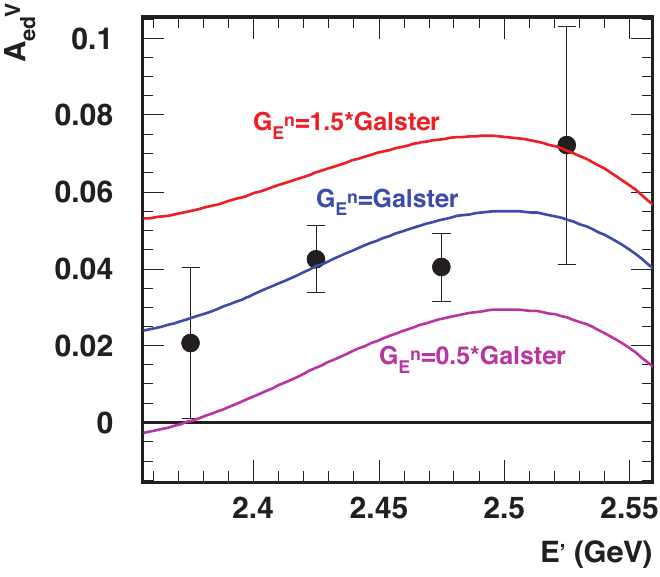}    
  \includegraphics[width=0.32\textwidth]{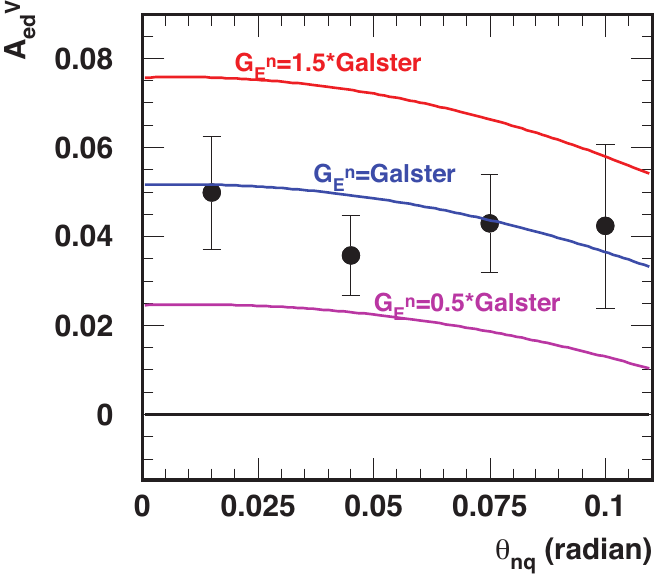}    
 \end{center}
  \caption[E93-026 Results of $A_{ed}^{V}$]{Double spin asymmetry $A_{ed}^{V}$ as measured in Hall C, vs. scattered electron energy ($E^\prime$) and vs. the angle between the neutron and the $\vec{q}$ vector.  The theoretical curves are predicted asymmetries using different scaled values of $G_{E}^{n}$~\cite{zhu2000measurement}.}
  \label{avedhallc:fig}
\end{figure}
Within Arenh\"{o}vel's model, all ingredients including isobar currents and relativistic corrections are needed to describe most of the BLAST data. In particular, a reasonable description of $A_d^T$ in parallel kinematics (top right panel of Fig.~\ref{fig:BLAST}) requires the inclusion of FSI effects at larger $p_m$.  This is in qualitative agreement with our findings (see below).  Figure~\ref{fig:BLAST} also shows that $A^V_{ed}$ is described rather well by the simpler PWBA out to significantly higher missing momentum than is the case for $A_d^T$.

\subsection{Hall C}
The E93-026 experiment in Hall C at the Thomas Jefferson National Accelerator Facility measured the neutron electric form factor $G^n_E$ in \deep~quasi-elastic scattering.   In this measurement~\cite{zhu2000measurement} the neutron was detected instead of the proton and $G^n_E$ was extracted by comparing the measured $A^V_{ed}$ to theoretical predictions by Arenh\"{o}vel using variations of the parameterization of $G^n_E$ by Galster \textit{et al.}~\cite{Galster:1971kv}.

The target used in this experiment was a \nd~target similar to the target used in the present study.  The experiment was limited to $Q^2$ = 0.5 (GeV/$c)^2$ and the missing momentum was less than 180 MeV.  The results are shown in Fig.~\ref{avedhallc:fig}.

\section{EXPERIMENTAL SETUP}\label{section:experiment}
This analysis is based on data from the EG1b group of experiments that took place at the Thomas Jefferson National Accelerator Facility (Jefferson Lab) located in Newport News, Virginia.  The Continuous Electron Beam Accelerator Facility (CEBAF) at Jefferson Lab provided polarized electron beams from 1.6 to 5.8 GeV energy to the CEBAF Large Acceptance Spectrometer (CLAS) in Jefferson Lab's Hall B.  The beam polarization was periodically measured with a M\o ller polarimeter.  The experimental run was conducted in 2000--2001 for a period of seven months.

\begin{figure}[hb!]
\centering
\includegraphics[width=0.4\textwidth]{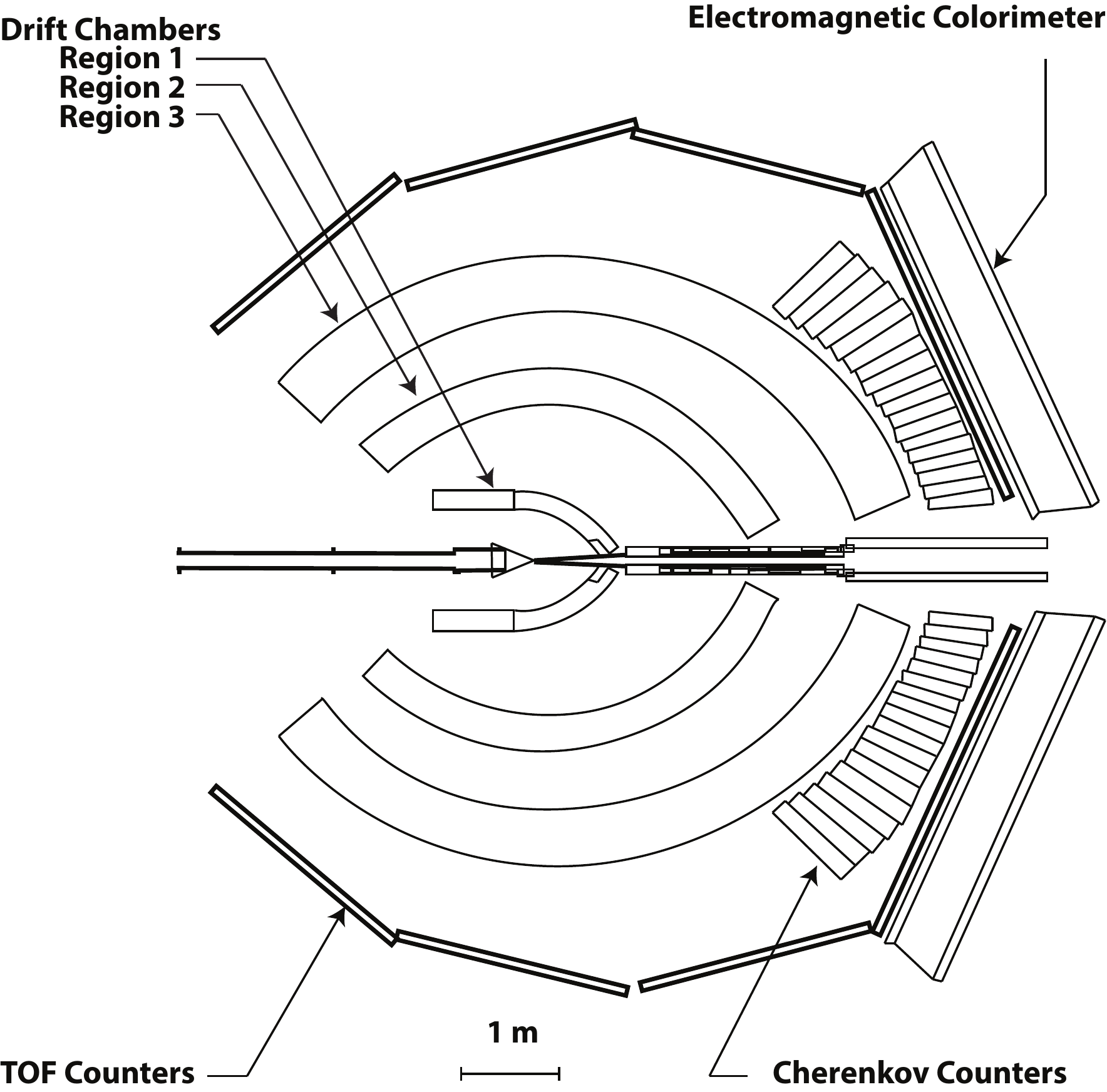}
\caption[The CLAS Detector]{A cross section view of the CLAS detector.}
\label{fig:clas}
\end{figure}

CLAS is divided by six superconducting coils into six symmetric sectors with several layers of  particle detectors.  The coils produce a mostly azimuthal magnetic field. There are three layers of drift chambers (DC) for tracking in this field, followed by a layer of scintillator counters (TOF) for time-of-flight  measurements.  Cherenkov counters and electromagnetic calorimeters in the forward regions are used to identify the scattered electrons.  A Faraday Cup is used to measure the total accumulated beam charge. The CLAS Data Acquisition (DAQ) system collected data at a 3-4 kHz event rate, triggered by a coincidence of the  signals above threshold from the electromagnetic calorimeters and Cherenkov counters.  A detailed description of CLAS and its systems can be found in~\cite{mecking2003cebaf}.

For the EG1b run, longitudinally polarized electrons were scattered from several different targets placed alternatively on the center line of CLAS and immersed in a liquid Helium bath at 1 K. These included longitudinally polarized proton (\nh) and deuterium (\nd) targets, as well as auxiliary $^{12}$C and liquid $^{4}$He (``empty'') targets.  Beam and target polarizations were either parallel or anti-parallel with respect to each other and the beam direction.  The two polarized targets, \nh~and \nd, were polarized by the DNP method.  The deuterium target maintained roughly 20\%-40\% polarization during data collection.  The polarization was measured in real time using a nuclear magnetic resonance (NMR) system; however, the final determination was based on measured double spin asymmetries, as explained below. To minimize depolarization of the targets due to heating and radiation damage, the electron beam was rastered over the surface of the targets in a spiral pattern during the experimental run. The targets were periodically annealed to remove extra paramagnetic radicals and restore polarization.  Further information on the polarized target can be found in~\cite{keith2003polarized}.

The EG1b group of experiments collected data using several different experimental configurations.  The polarized electron beam had energies of 1.606, 1.723, 2.286, 2.561, 4.238, 5.615, 5.725, and 5.743 GeV with current from 0.3 nA to 10 nA.   The current of the main toroidal magnet was set at 2250 A or 1500 A and was switched from positive to negative polarity at times. Positive current polarity resulted in electrons being bent towards the beam axis (inbending), while negative polarity led to outbending electrons and extended the accepted kinematics to lower scattering angles. All EG1b run sets with usable electron beam data are labeled by beam energy and torus polarity (\textit{e.g.}, a 4.2 GeV electron beam run with a positive torus current is labeled as 4.2+) and are listed in Table \ref{table:run}. These set labels are used throughout this paper. The 2.3 GeV data had too few events for the present analysis to yield statistically significant results and was therefore not included. The remaining beam energies were combined into four groups, with average (nominal) energies of 1.6, 2.5, 4.2, and 5.7 GeV. 

\begin{table}[htbp] 
\small  \centering
\caption[EG1b ND$_{3}$ Run Sets by Beam Energy and Torus Current]
{All ND$_{3}$ EG1b run sets with usable electron beam data, organized by beam energy and torus polarity. These set labels are used throughout this paper.  The sets were further distinguished by the polarization of the target during each run. Note that the 2.3+ set was ultimately not used in the present analysis. The remaining sets were combined into four major groups, as indicated by the horizontal lines.}
\vspace{1cm}
\begin{tabular}{cccc}
\hline
\hline
Set Label  &  E$_{Beam}$(GeV) & I$_{Torus}$(A) \\
\hline
1.6+  &  1.606  & +1500\\
1.6$-$   &  1.606  & $-$1500\\
1.7+   &  1.724  & +1500\\
1.7$-$  &  1.724  & $-$1500\\
\hline
[ 2.3+  &  2.288  &  +1500 ] \\
2.5+   &  2.562  &  +1500\\
2.5$-$   &  2.562  &  $-$1500\\
\hline
4.2+   &  4.239  &  +2250\\
4.2$-$   &  4.239  &  $-$2250\\
\hline
5.6+   &  5.627  &  +2250\\
5.6$-$   &  5.627  &  $-$2250\\
5.7+   &  5.735  &  +2250\\
5.7$-$   &  5.735  &  $-$2250\\
5.7$-$   &  5.764  &  $-$2250\\
\hline
\hline
\end{tabular}
\label{table:run}
\end{table}

Additional experimental information can be found in the archival publications~\cite{Fersch2015aa,Guler:2015hsw} for EG1b.

%%%%%%%%%%%%%%%%%%%%%%%%%%%%%%%%%%%%%%%%%%%%%%%%%

\section{ANALYSIS}\label{section:analysis}
\subsection{Data Selection}
The analysis presented here builds on the previously published standard analysis of the complete EG1b data set~\cite{Guler:2015hsw,Fersch2015aa}, including all calibrations, corrections, basic cuts and quality checks.  From that analysis, we selected reconstructed events containing an electron and either only one proton or one proton and one neutral particle (which could be the recoiling neutron). Electrons were identified through cuts on the signals in the Cherenkov counters and electromagnetic calorimeters, while protons were selected based on their time of flight (measured with the TOF) and their momentum. Fiducial cuts on both electrons (to exclude regions of rapidly varying detection efficiency) and protons (to avoid both the CLAS torus and polarized target coil enclosures) were applied. All further details can be found in ~\cite{Guler:2015hsw,Fersch2015aa}.

\begin{figure}
\centering
\includegraphics[width=0.5\textwidth]{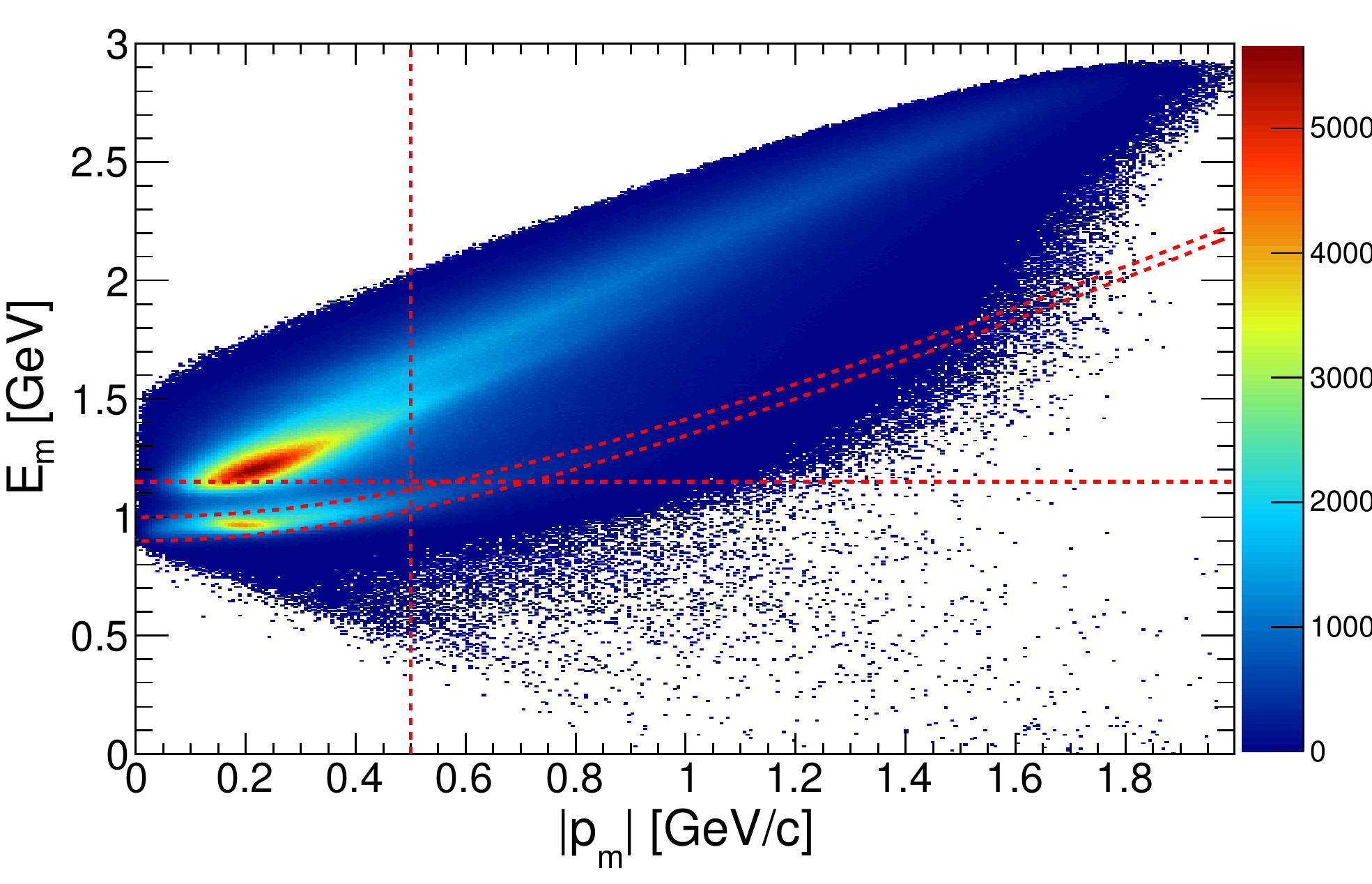}
\caption[]{A plot of missing energy vs. missing momentum where the red lines represent the cuts placed on the data.}
\label{fig:MMvsEM}
\end{figure}

Information on the neutron kinematics was gained through use of the conservation of energy and momentum.  The missing energy ($E_{m}$) was calculated under the assumption that the reaction took place on a deuteron at rest, as
\begin{equation}
E_{m} = m_{d} +\nu - E_{p},
\label{eq:mE}
\end{equation}
where $m_{d}$ is the mass of the deuteron, $\nu = E - E^\prime$ is the energy of the virtual photon,
and $E_p$ is the measured energy of the proton.  A cut $E_{m}<1.15$ GeV was applied to reduce the size of the data sample, to include only events of interest.  The missing momentum ($p_{m}$) was calculated as
\begin{equation}
\vec{p}_m=\vec{q}-\vec{p}_p,
\end{equation}
where the 3-momentum of the virtual photon, $\vec{q}$ , is calculated from the measured electron kinematics
%$\vec{q}$ is the 3-momentum of the virtual photon (the momentum transfer from the scattered electron) and 
and $\vec{p}_p$ is the momentum of the detected proton.  Because the nuclear background overwhelms the signal
from the deuteron
 at high missing momenta, we only analyzed events  with $| \vec{p}_{m} | < 0.5$ GeV$/c$. %where the count rate was higher than the background.  
Finally, the missing mass ($M_{m}$) was calculated as
\begin{equation}
 M_m=\sqrt{E_m^2-p_m^2} .
 \label{eq:MM}
\end{equation}
Examples of missing mass distributions for different kinematic bins can be found in Section~\ref{sec:inelas}.  The mass of the neutron is known to be 0.94 GeV/c$^2$ and a missing mass cut was implemented to remove multi-particle final states:
\begin{equation}
0.9<M_{m}<1.0\mbox{~(GeV/c$^{2}$).} 
\label{eq:mmcut}
\end{equation}
The combination of these cuts can be seen in Fig. \ref{fig:MMvsEM}. The curved lines identify the missing mass cut selecting exclusive $pn$ final states from a deuteron.

\subsection{Binning of Data}
\label{binning:sec}
The asymmetries presented in this paper were calculated as a function of three kinematic variables: the squared four-momentum transfer, $Q^{2}$, the cosine of the angle between the virtual photon and the neutron momenta, $\cos\theta_{nq}$, and the missing momentum, $p_{m}$.  We integrated over the angle $\phi$ between the leptonic and hadronic plane.

For compatibility with the main EG1b analysis, we combined several of the standard EG1b $Q^{2}$ bins~\cite{Guler:2015hsw,Fersch2015aa} into four larger bins in the range  $0.131-3.17$ (GeV/$c)^2$ (smaller bins would have yielded too limited statistics in the quasi-elastic region).  Table \ref{Q2:table} shows these four $Q^{2}$ bins. 
\begin{table}[htbp]  
 \centering
\small
\caption[]{$Q^2$ bins used in this analysis}
\vspace{0.5cm}
\begin{tabular}{ccc}
\hline
\hline
Bin  &  $Q^2_{min}$ (GeV/c)$^2$ &  $Q^2_{max}$ (GeV/c)$^2$\\
\hline
0 & 0.131 & 0.379 \\
1 & 0.379 & 0.770 \\
2 & 0.770 & 1.56 \\
3 & 1.56 & 3.17 \\
\hline
\hline
\end{tabular}
\label{Q2:table}
\end{table}

The data were divided further into three regions of $\cos\theta_{nq}$ from -1.0 to 1.0, corresponding to the spectator neutron moving backwards, sideways or forward relative to $\vec{q}$. The exact ranges of these $\cos\theta_{nq}$ bins are shown in Table \ref{costhetanq:table}.
\begin{table}[htbp] 
\centering
\caption[$\cos\theta_{nq}$ Bins]{The $\cos\theta_{nq}$ bins used in this analysis.}
\vspace{1cm}
\begin{tabular}{ccc}
\hline
\hline
Bin & $\cos\theta_{nq}^{min}$ & $\cos\theta_{nq}^{max}$ \\
\hline
0 & -1.0 & -0.35 \\
1 & -0.35 & 0.35\\
2 & 0.35 & 1.0\\
\hline
\hline
\end{tabular}
\label{costhetanq:table}
\end{table}

The final binning is in missing momentum.   We are interested in missing momenta ranging from 0.0 to 0.5 GeV$/c$.  This range was divided into five missing momentum bins shown in Table \ref{pmiss:table}.  
\begin{table}[htbp] 
\centering
\caption[Missing Momentum Bins]{The missing momentum bins used in this analysis.}
\vspace{1cm}
\begin{tabular}{ccc}
\hline
\hline
Bin & $p_m^{min}$ (GeV$/c$) & $p_m^{max}$ (GeV$/c$)\\
\hline
0 & 0.00 & 0.05 \\
1 & 0.05 & 0.15\\
2 & 0.15 & 0.25\\
3 & 0.25 & 0.35\\
4 & 0.35 & 0.50\\
\hline
\hline
\end{tabular}
\label{pmiss:table}
\end{table}

In total, we have 60 bins for each of our 4 major beam energy groups listed in Sec.~\ref{section:experiment}.  Because all of our (three-dimensional) bins are rather wide, any comparison with theoretical calculations requires the latter to be integrated over the same bins, weighted with the distribution of actually observed events over each bin.  In our final results, we present only asymmetries for those bin and beam energy combinations where the following conditions were fulfilled:
\begin{enumerate}
\item The missing mass distribution covers the full region of our cut, $0.9 < M_m < 1$ GeV/c$^{2}$, and shows a clear, distinct peak inside that region.
\item The difference between measured \nd\  counts and inferred background counts from non-deuterium components of the target (see below) exceeded two standard deviations above zero.
\end{enumerate}
These criteria are further explained in the following section.

\subsection{Determination of the Double Spin Asymmetries}
For each of the bins defined above, the raw asymmetry was calculated as
\begin{equation}\label{eq:Araw}
A_{raw}=\frac{n^{+}-n^{-}}{n^{+}+n^{-}} , 
\end{equation}
where $n^{\pm}$ is the normalized count per helicity state and is defined as
\begin{equation}
n^{-} = \frac{N^{\upuparrows}}{FC^{\upuparrows}} \mbox{~~~~and~~~~} n^{+} = \frac{N^{\uparrow\downarrow}}{FC^{\uparrow\downarrow}} , 
\end{equation}
where $FC$ is the Faraday cup integrated charge.  The arrows indicate parallel and anti-parallel
beam and target polarization. The Faraday cup signal was gated on the DAQ live-time to correct for dead time effects.

In the following, we discuss all corrections that had to be applied to extract the final physics
asymmetries.

\subsubsection{Inelastic Background}
\label{sec:inelas}

\begin{figure} [ht!]
%\centering
\includegraphics[width=0.23\textwidth]{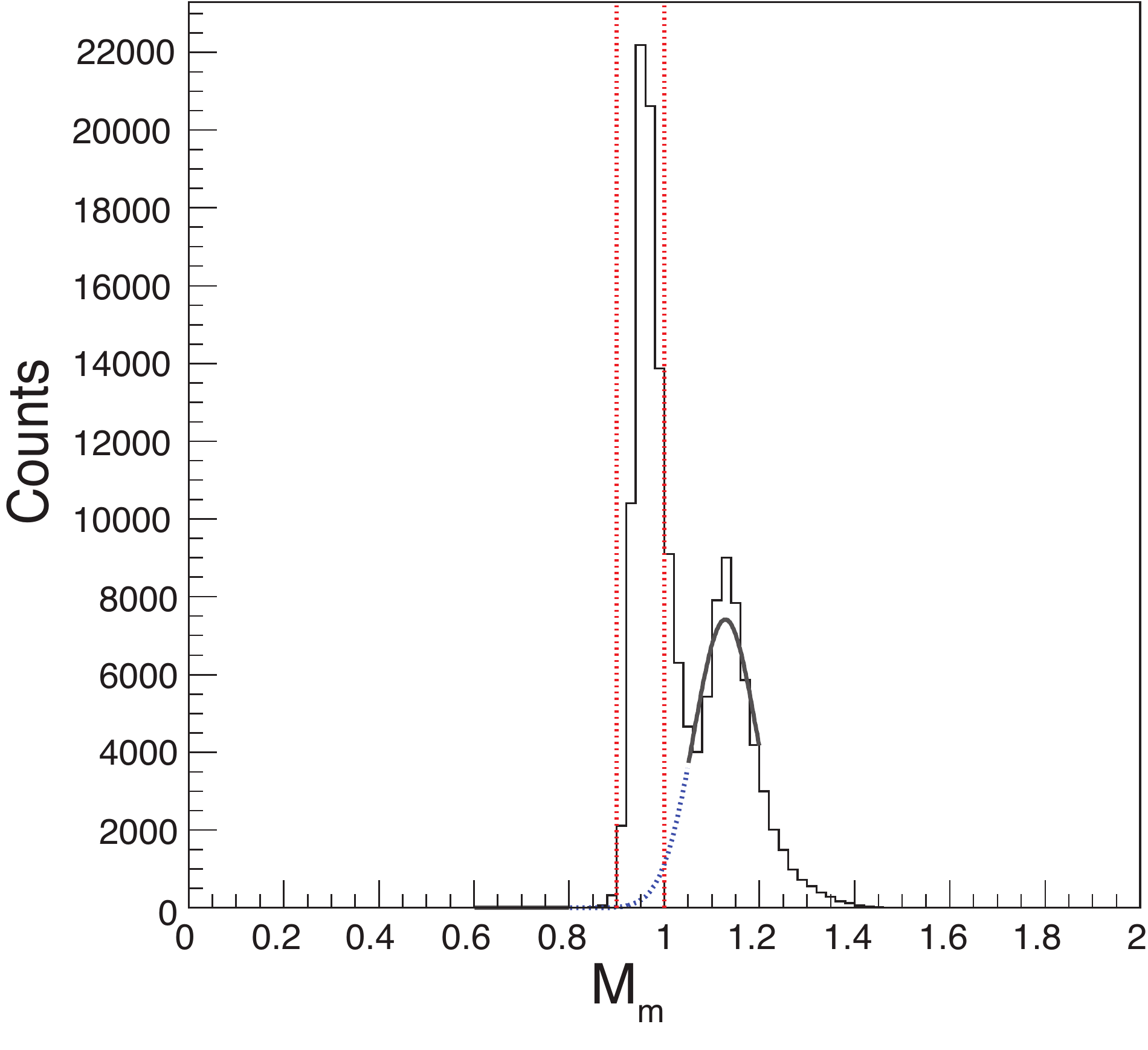}
\includegraphics[width=0.23\textwidth]{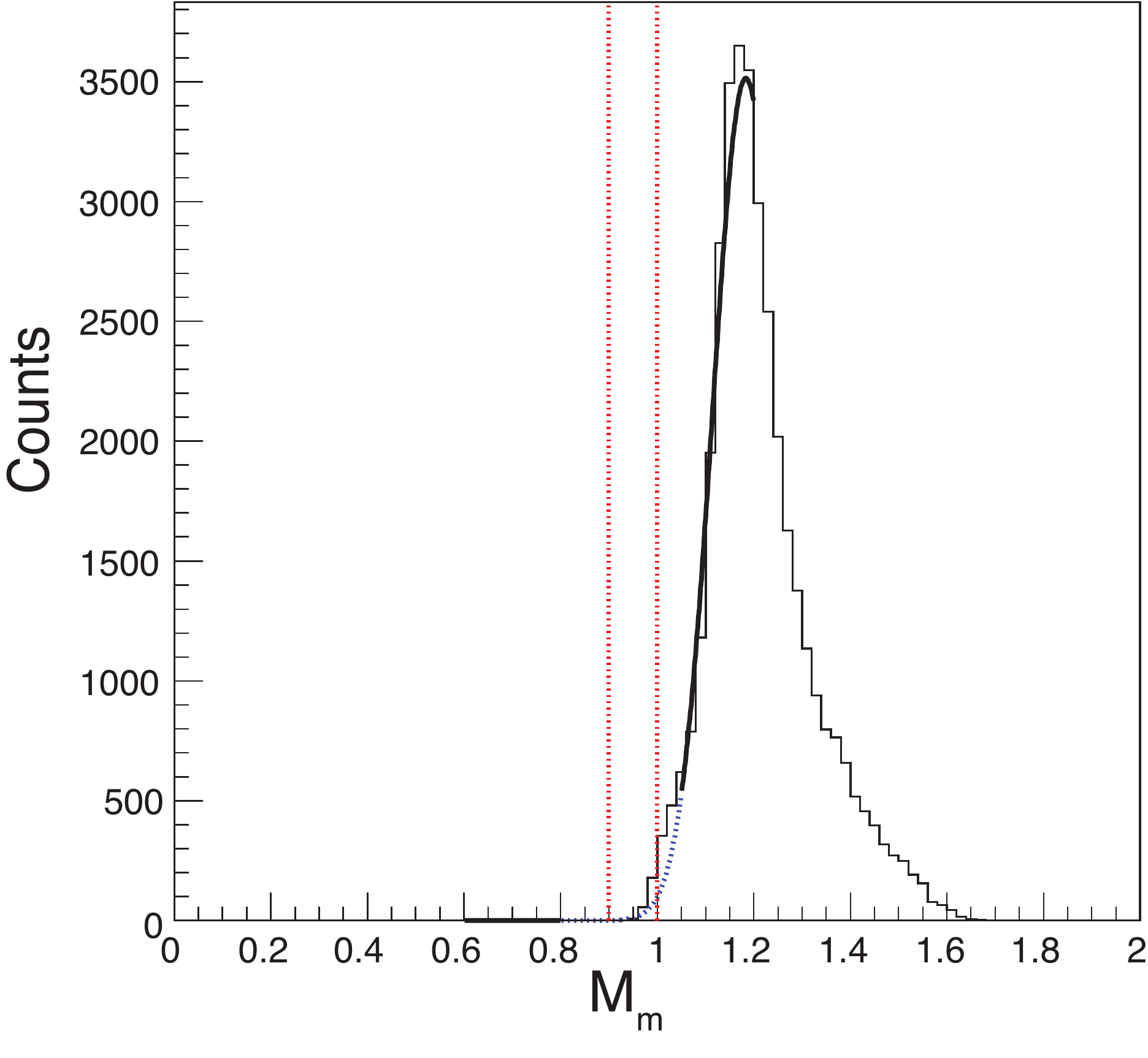}
\caption[]{Distribution of events in missing mass (Eq.~(\ref{eq:MM})) for two different kinematic bins (Left: $E_{Beam} = 1.6$ GeV, $Q^2$ bin 1, $p_m$ bin 1, $\cos\theta_{nq}$ bin 1; Right: $E_{Beam} = 1.6$ GeV, $Q^2$ bin 1, $p_m$ bin 1, $\cos\theta_{nq}$ bin 2). The inelastic background was fitted by a Gaussian tail shown as the solid and dotted black lines (the dotted line is the interpolation between the two fit regions). The right hand plot is an example for a kinematic setting where, due to CLAS acceptance, no peak is seen within the missing mass cut Eq.~(\ref{eq:mmcut}), indicated by vertical dotted red lines. Bins such as these were discarded in the further analysis.}
\label{fig:MM1}
\end{figure}

\begin{figure} [h!]
%\centering
\includegraphics[width=0.23\textwidth]{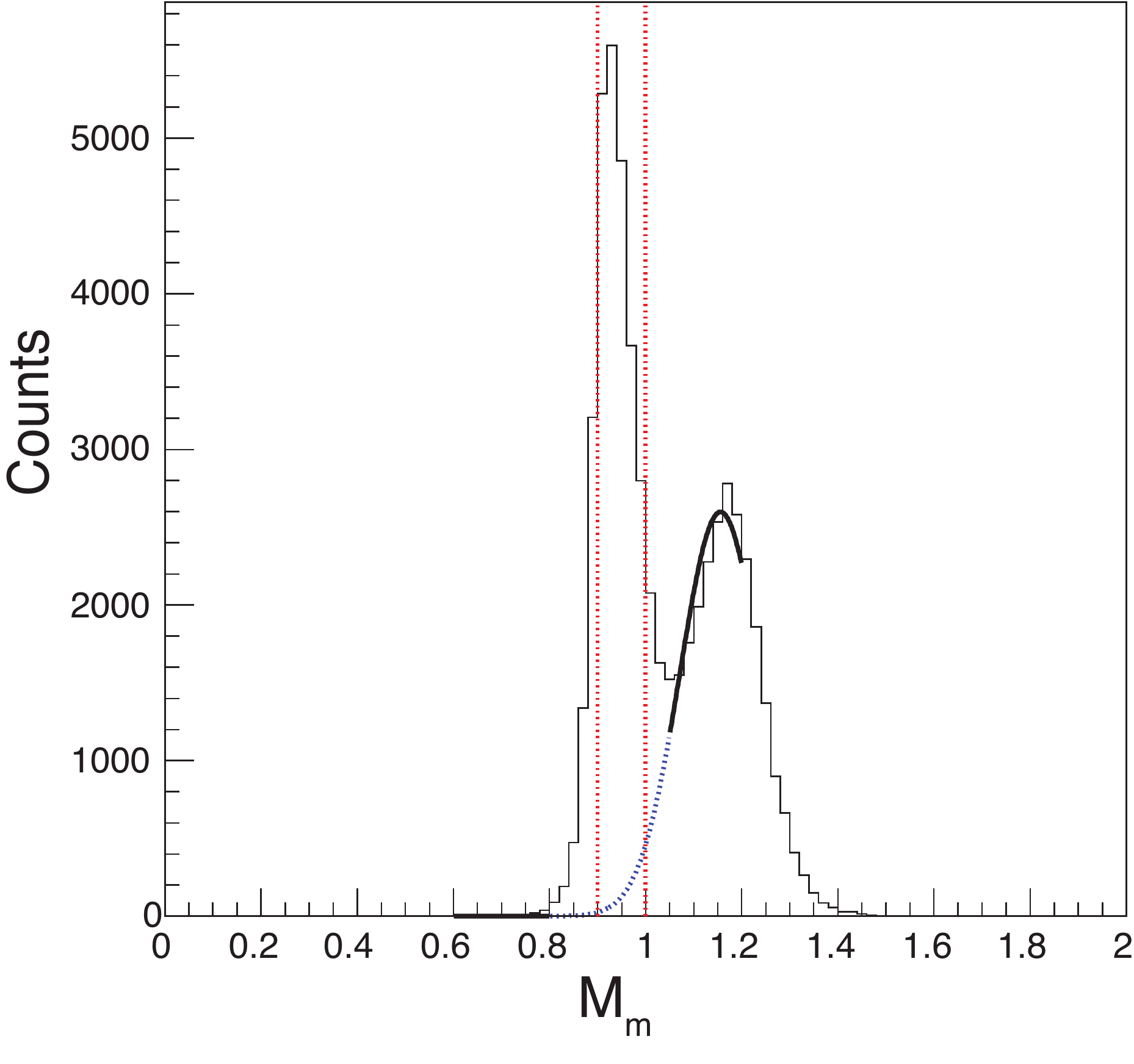}
\includegraphics[width=0.23\textwidth]{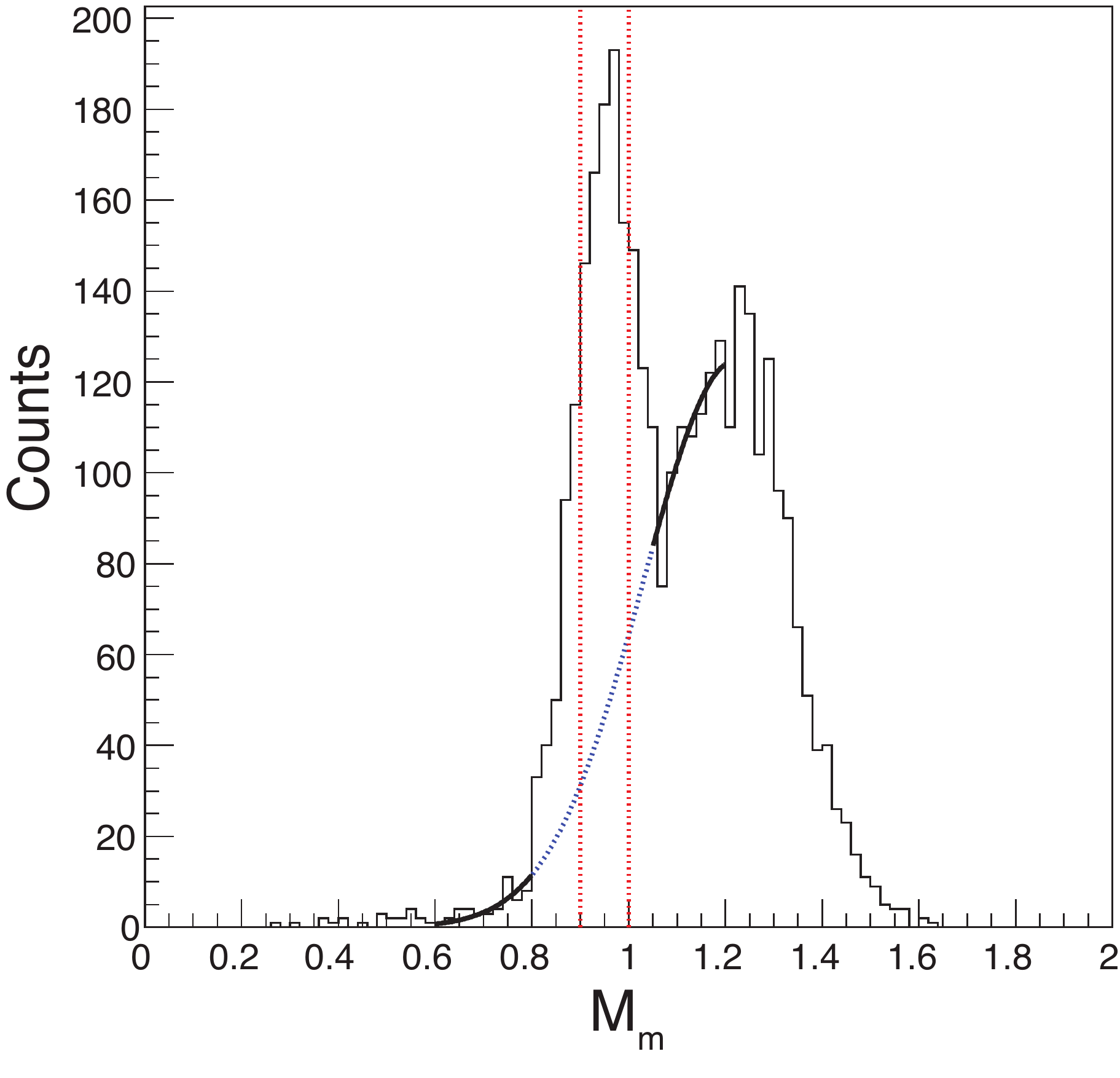}
\caption[]{Same as Fig.~\ref{fig:MM1} for two additional kinematic bins (Left: $E_{Beam} = 4.2$ GeV, $Q^2$
bin 1, $p_m$ bin 3, $\cos\theta_{nq}$ bin 0; Right: $E_{Beam} = 5.7$ GeV, $Q^2$ bin 3, $p_m$ bin 4, $\cos\theta_{nq}$ bin 1).}
\label{fig:MM2}
\end{figure}

Due to finite detector resolution,
a small fraction of inelastic events (with additional particles in the final state) could be present within the region of our missing mass cut, see Figs.~\ref{fig:MM1}--\ref{fig:MM2}.  This background was studied in great detail to correct the extracted asymmetries for this contribution (for this study, we removed the cut on $E_{m}$, Eq.~(\ref{eq:mE})).  We determined the fraction of counts $f_{back} = n_{back}/n_{total}$ from such inelastic events by simultaneously fitting the missing mass distribution for every kinematic bin and for every beam energy, torus polarity, and target polarization on both sides of the elastic peak, covering the range $0.6<M_m<0.8$ GeV/c$^{2}$ and $1.1<M_m<1.2$ GeV/c$^{2}$. The two regions were used to account for background tails from the inelastic region, $M_m > 1.07$ GeV/c$^{2}$, that could extend to lower $M_m$ regions due to kinematic smearing. We found that a Gaussian tail provided a good fit in all cases (black solid and dotted line in Figs.~\ref{fig:MM1}--\ref{fig:MM2}). This fit was then integrated over our missing mass cut to estimate $n_{back}$.
 
Simultaneously, the count rate asymmetry in the upper missing mass region, $1.1<M_m<1.2$ GeV/c$^{2}$, was used to estimate the asymmetry, $A_{back}$, of this background. The measured asymmetry was then corrected to get the quasi-elastic asymmetry only:
\begin{equation}\label{eq:Aqe}
A_{qe}=\frac{A_{raw} - A_{back} f_{back}}{1 - f_{back}} .
\end{equation}

This correction changed the final physics asymmetries by typically less than 10\% of their values, and much less than their statistical uncertainties. We use this change in the asymmetries as a generous upper limit on the systematic uncertainty for this correction. 

\subsubsection{Unpolarized Background Corrections}
\label{df:sec}

The denominator in Eq.~(\ref{eq:Araw}) contains counts not only from the (desired) polarized deuterium nuclei, but also all other components of the target (including the nitrogen in the \nd~molecules and the liquid $^4$He coolant as well as various window foils). Since these target components are unpolarized, they do not affect the numerator; see, however, Section~\ref{polcor:sec}. After determining the contribution from this unpolarized background, $n_{A-D}$, to both $n^+$ and $n^-$ in Eq.~(\ref{eq:Araw}), the undiluted asymmetry can be extracted as follows:
\begin{equation}\label{eq:undiluted}
A_{undil}=\frac{n^{+}-n^{-}}{n^{+}+n^{-}-2n_{A-D}} = \frac{n^{+}-n^{-}}{n^{+}+n^{-}-n_B} ,
\end{equation}
where $n_B=2n_{A-D}$.
We further define the dilution factor as
\begin{equation}\label{eq:dilutionfactor}
F_{D}=\frac{n^{+}+n^{-}-n_B}{n^{+}+n^{-}}=1.0-\frac{n_B}{n^{+}+n^{-}} .
\end{equation}
The raw asymmetry can then be corrected for the unpolarized background by dividing out the
dilution factor giving the equation for the undiluted asymmetry as
\begin{equation}\label{eq:Adil}
A_{undil}=\frac{A_{raw}}{F_{D}}.
\end{equation}

To calculate $n_{A-D}$ (or, equivalently, the dilution factor), we modeled the contribution from unpolarized target components as a combination of counts from auxiliary measurements on two additional target cells: a cell containing only a disk of $^{12}$C (``C'') and another cell without target material (``MT''), both immersed in the same liquid $^4$He bath. After normalizing these counts to the integrated Faraday cup and accounting for the thickness of all components for each target, we could extract a ``pure'' carbon target count rate $n^{\prime}_{C}$, and a ``pure'' helium target count rate $n^{\prime}_{He}$, from these auxiliary measurements. The unpolarized background was then  calculated as
\begin{equation}\label{eq:nad}
n_{A-D}=n_{MT}+l_{A}(\frac{\rho_{A}}{\rho_{C}l_{C}}\frac{7}{6}n'_{C}-n'_{He})
\end{equation}
for each of our kinematic bins.  Here, $n_{MT}$ is the count rate on the ``MT'' target, $\ell_{A}$ is the packing fraction (the equivalent length of the target cell after accounting for the percentage of its volume occupied by ammonia beads), and $\ell_{A}\rho_{A}/\rho_{C}l_{C}$ is the relative thickness (in target atoms per cm$^2$) of the ammonia {\it vs.} the carbon target. The factor 7/6 accounts for the fact that there are 7 protons in $^{15}$N vs. 6 in $^{12}$C that could partake in  quasi-elastic $(e,e^\prime p)$ knockout. Finally, the term $l_{A} n'_{He}$ subtracts the amount of $^4$He liquid displaced by the ammonia from the ``MT'' target.

The archival EG1b papers~\cite{Guler:2015hsw,Fersch2015aa} explain how each of the parameters
entering Eq.~(\ref{eq:nad}) was determined. We varied all parameters within their uncertainties to estimate the possible spread of the magnitude of the unpolarized background and its effect on the extracted asymmetries (with a resulting systematic uncertainty for the latter between 4\% and 11\% of their nominal values). 

\subsubsection{Beam and Target Polarization}\label{sec:pbpt}
In addition to the dilution by unpolarized target components, the measured asymmetry must also be corrected for the target and beam polarization,
\begin{equation}\label{eq:Apolnorm}
A_{||}=\frac{A_{undil}}{P_b P_t}.
\end{equation}
In principle, both these quantities were measured either continuously (target polarization, through NMR) or at regular intervals (through asymmetry measurements in M\o ller scattering).
However, the target material can undergo local depolarization due to radiation damage and heating from exposure to the electron beam, rendering NMR measurements somewhat unreliable.  Instead, the product of the beam and target polarization ($P_b P_t$) was determined directly from the data.  The values used in this analysis were obtained from Ref.~\cite{Guler:2015hsw}.  In that work, values of $P_b P_t$ were extracted from the EG1b data set by comparing a theoretical value of $A_{||}$ to a background-corrected measurement of $A_{||}$ doe quasi-elastic scattering off the deuteron. We used both the extracted values and the estimated uncertainty on $P_b P_t$ from Ref.~\cite{Guler:2015hsw} to estimate the systematic uncertainties of our final results due to this source.

\subsubsection{Target Contamination}
\label{tarcont:sec}
In addition to unpolarized nucleons in the target, we must also correct our asymmetries for the potential presence of other polarized nucleons outside deuterium. Experience has shown that solid polarized $^2$H targets typically contain small amounts of polarized materials other than $^2$H. However, a more recent experiment in CLAS with a deuterated ammonia target found a surprisingly large contribution from polarized free protons to the measured asymmetry~\cite{Prok:2014ltt}.  Therefore, we performed a careful study of the EG1b target, using the method from Ref.~\cite{Prok:2014ltt}, to identify any such polarized proton contamination that would affect the results of this analysis.

The method used in this analysis relies on a comparison of exclusive $e-p$ elastic events from the proton and quasi-elastic events on the deuteron.  (Quasi-)elastic events on \nh, \nd, and $^{12}$C targets were selected by applying the particle identification cuts with an additional cut of
\begin{equation}
||\phi_{e}-\phi_{p}|-180.0\degree| < 3.0\degree,
\label{eq:targconcut}
\end{equation}
where $\phi_{e}$ is the azimuthal angle of the scattered electron and $\phi_{p}$ is the azimuthal angle of the scattered proton.

The CLAS detector is much more precise at determining polar angles than momenta for detected particles.  Using the polar component of the proton's momentum ($p_{\theta}$), we can separate quasi-elastic events on the deuteron and heavier nuclei and elastic events on the proton.  The difference between the measured and expected polar component of the proton's momentum was calculated as
\begin{equation}
\label{fermi:eq}
\Delta p_{\theta}=|p_{p}|(\sin(\theta_{p})-\sin(\theta_{q})),
\end{equation}
where $p_{p}$ is the momentum of the proton, $\theta_{p}$ is the polar angle of the proton, and $\theta_{q}$ is the polar angle of the
virtual photon.  For elastic scattering, this quantity is given by 
\begin{equation}
\label{thetaq:eq}
\tan(\theta_{q}) = \frac{1}{\left(\frac{E}{m_{p}}+1.0\right) \tan\left(\frac{\theta_{e}}{2.0}\right)},
\end{equation} 
where $E$ is the energy of the incoming electron beam and $m_{p}$ is the mass of the proton.  We used the relationship for elastic scattering to get the sharpest possible peak for $^1$H$(e,e^{\prime}p)$.  For quasi-elastic scattering, a broader peak is expected due to Fermi motion.  

\begin{figure}[htb!]
\centering
\includegraphics[width=0.5\textwidth]{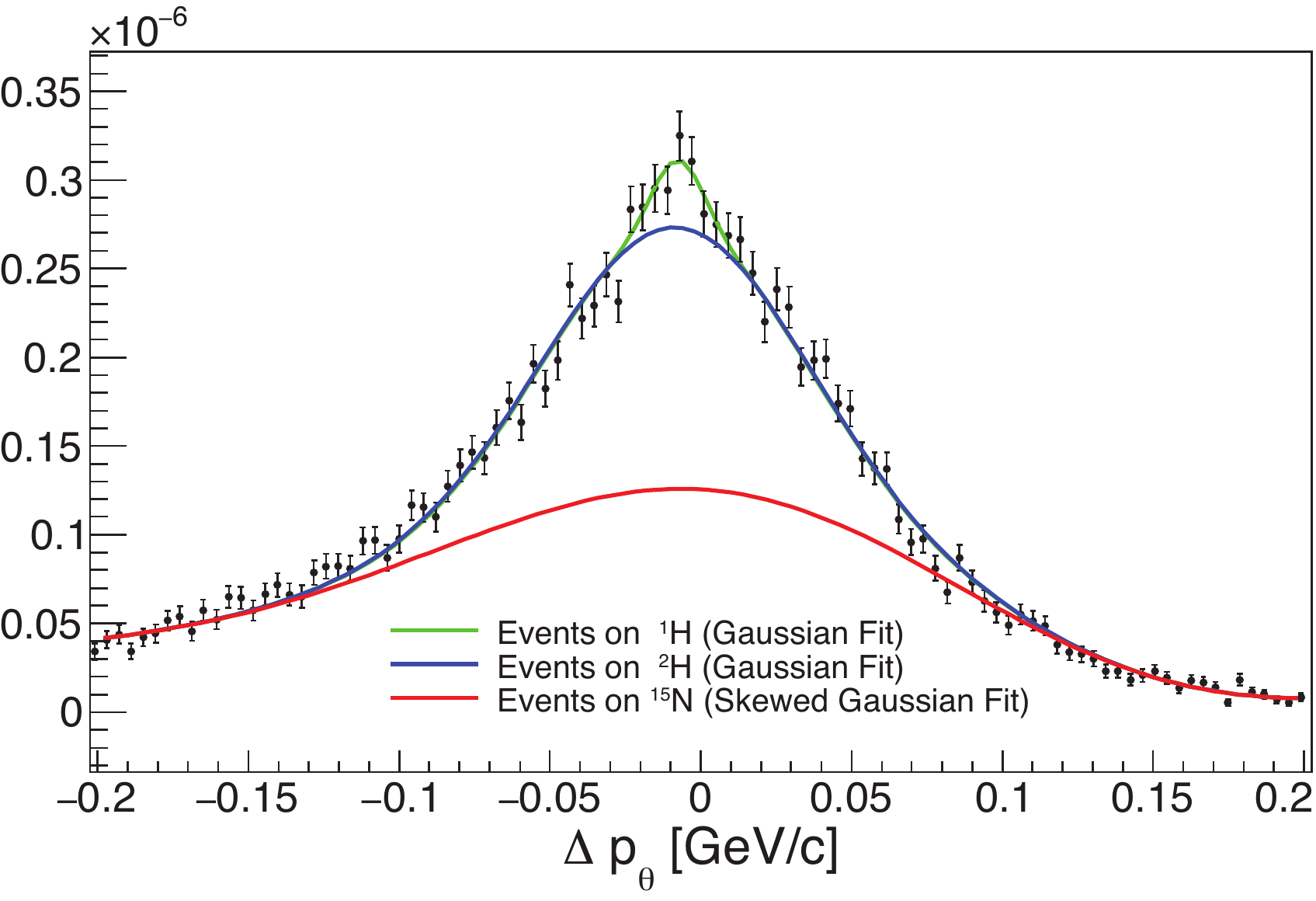}
\caption[$\Delta p_{\theta}$ for ND$_{3}$]{Distribution of counts vs. $\Delta p_{\theta}$ for ND$_{3}$ targets.  The $^1$H peak (green) can be seen on top of the $^2$H peak (blue) with the scaled background (red).} \label{results1:fig}
\end{figure}

First, we binned the data for the three targets in $\Delta p_{\theta}$, using the 4.2 GeV in-bending runs.  The count rates were normalized by the corresponding Faraday Cup counts.  We used a fit to the carbon target data to emulate the background from $^{15}$N and $^{4}$He in the ammonia targets.  The fit has a functional form with five parameters that were optimized for minimum $\chi^{2}$ in the region around $\Delta p_{\theta} = 0$.  The NH$_3$ data were fitted next as a sum of this (appropriately scaled) background and a narrow Gaussian centered at $\Delta p_{\theta} = 0$ for elastic scattering off $^1$H. 
Finally, keeping all fit parameters (other than the adjustable normalization factors) fixed for both the background and the free proton peak from the ~\nh~data, we fit the \nd~data by adding a second quasi-elastic (deuteron) peak to the other two contributions.  The results can be seen in Fig. \ref{results1:fig}.  The relative free proton contamination is then the ratio of the areas under the $^1$H and $^2$H peaks, corrected for the  suppression of quasi-elastic events on the deuteron due to the $\Delta \phi$ cut (Eq.~(\ref{eq:targconcut})). We find a contamination around 3.5\%.  This contamination may come from NH$_{3}$ impurities, frozen H$_{2}$O, or other sources.  The typical value used in previous analyses (EG1a~\cite{PhysRevC.67.055204}, E155~\cite{Anthony1999339}) is around 1.5\%, based on typical isotopic purities of \nd. 

\begin{figure}[htb!]
\centering
\includegraphics[width=0.5\textwidth]{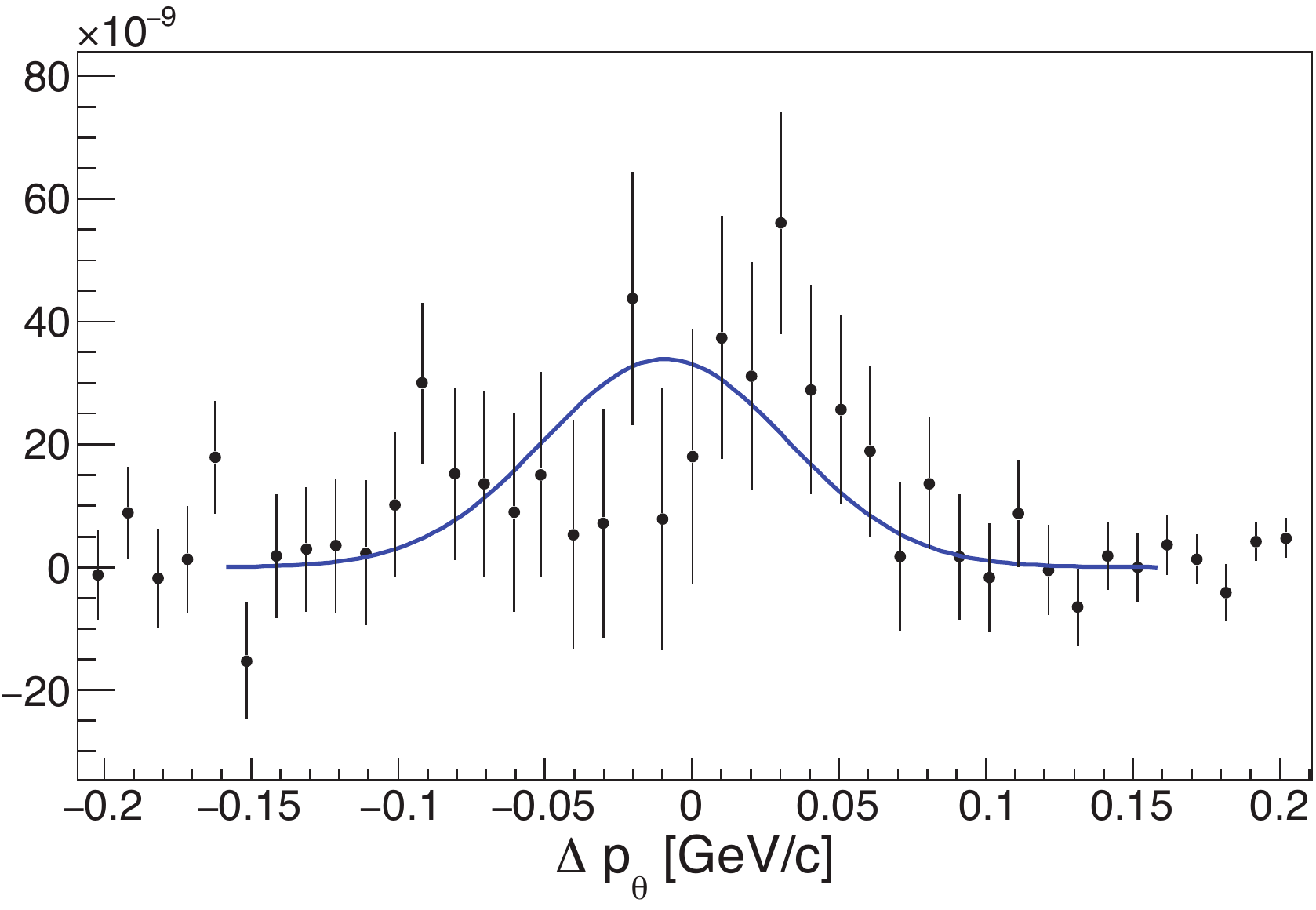}
\caption[$\Delta n$ for ND$_{3}$]{Count rate differences for antiparallel vs. parallel beam and target polarization for the ND$_{3}$ target, vs. $\Delta$p$_{\theta}$.  The distribution is fit with a free scale parameter for both D and H. The lowest $\chi^2$ results from a fit with no contribution from free protons. }\label{results3:fig}
\end{figure}

To determine to what degree this $^1$H contamination of the \nd~target was polarized, we used the difference between the normalized count rates for the two helicity states, $\Delta n = n^+ - n^-$. Contributions from unpolarized target components drop out in this difference. For the proton target, we see indeed a narrow peak without any background.  The corresponding distribution for the \nd~target (Fig. \ref{results3:fig}) shows only the broader deuteron peak; a fit with a double Gaussian (as before) yields zero as the most likely contribution from the narrow proton peak. This would indicate that (most of) the hydrogen contamination in the \nd~target is not polarized ({\it e.g.}, it could be due to frozen water contamination).  We can put an upper limit on the contamination from polarized protons of  2\% (one standard deviation), based on our fit.  Corrections due to this possible contamination are discussed in the next section.

\subsubsection{Polarized Background}\label{polcor:sec}
As stated previously, there are potentially  polarized nucleons outside of deuterium in the \nd~target, whose contribution to the measured asymmetry must be corrected for (the dilution factor only accounts for unpolarized background). The first source of such spin-dependent background stems from  bound protons in the $^{15}$N nuclei in the \nd~target that can become partially polarized from the DNP process.  (While approximately 2\% of the nitrogen nuclei are actually  $^{14}$N, they add only a negligible contribution to the measured asymmetry).  Finally, there are possibly polarized free protons, as discussed in the preceding section.
 
 The general formalism and specific assumptions entering these corrections on the asymmetry are discussed in the archival deuteron paper~\cite{Guler:2015hsw}.  In particular, this paper shows that the correction is of the general form
 \begin{equation}
\label{PolNitCorr:eqn}
A_{||}^{corr} = C_{1}\left(A_{||} - C_{2}A_{p}\right),
\end{equation}
and discusses the individual contributions to the coefficients $C_{1}$ and $C_{2}$.

 In the context of quasi-elastic scattering on the proton with small missing momenta (our first two $p_m$ bins, 0 and 1), we can make the simplifying assumption that all ``false'' asymmetries are proportional to the proton asymmetry $A_{p}$ alone, as is the measured asymmetry $A_{||}$. Hence, the correction becomes a simple multiplicative factor: 
 \begin{equation}
\label{PolNitCorr:eqn2}
A_{||}^{corr} = C_{q.e.} A_{||} .
\end{equation}
This factor depends on the kinematical bin, and is composed of three components:
\begin{enumerate}
\item 
The measured asymmetry is {\em reduced} relative to expectations if some of the deuteron atoms  are replaced by unpolarized hydrogen (e.g., in the form of H$_2$O molecules replacing some \nd~ones). From our discussion in the previous section, we assume that this is at most a 4\% effect. 
\item 
On the other hand, we cannot exclude a contribution from free protons that are at least partially polarized. This would increase the measured asymmetry and require an opposite correction of roughly the same magnitude, following our discussion in the previous section. 
\item
Finally, bound protons inside $^{15}$N can also be partially polarized. In a simple shell-model, one of the 7 protons occupies an unpaired $1p_{3/2}$ orbit, carrying a polarization of roughly -1/3 relative to the overall nuclear polarization ($P_{^{15}N}/P_D \approx 0.4 - 0.5$). This latter contribution is further suppressed by the larger Fermi momentum of bound protons in nitrogen as opposed to deuterium; in fact, it is proportional to the unpolarized background in a given kinematic bin. 
\end{enumerate}
For these reasons, we can write the combined effect of all of these corrections as
\begin{eqnarray}
A^{corr}_{||} = \frac{1}{P_b P_t} \frac{n^+ - n^-}{a(n^+ + n^-) - b n_B } 
\end{eqnarray}
(compare with Eq.~(\ref{eq:undiluted})). 
A careful study found that $a$ falls somewhere between $a = 0.976$ and $a = 1.015$, while $b$ is in the interval $0.97 \le b \le 1.004$. We estimated the systematic uncertainty of the final results resulting from this correction by varying both $a$ and $b$ within these limits.

For the highest $p_m$ bins (bins 2, 3, and 4),  free protons do not contribute, but the bound protons from $^{15}$N may have a different asymmetry than the bound proton in deuterium.  Unlike the lower $p_m$ bins, we therefore do not assume that  the bound proton asymmetry is proportional to the measured asymmetry.  Hence, we use Eq. (\ref{PolNitCorr:eqn}) where $C_{1} = 1$ and $C_{2}A_{p}$ corrects for the contribution from bound protons. It is once again proportional to the unpolarized background in each bin,
\begin{align}
 C_2  \approx 0.011 \frac{n_B}{n_+ + n_- - n_B} \approx0.03\mbox{ to } 0.18 ,
\end{align}
where the factor 0.011 accounts for the relative number and polarization of protons bound in nitrogen vs. deuterium.  The variation in $C_2$ corresponds to increasing missing momenta, where protons bound in nitrogen play a bigger role.  The high end for $C_2$ is an extreme value that applies only for the highest $p_{m}$ bin, where other statistical and systematic uncertainties are still larger. Meanwhile, the values for $A_{p}$ in Eq. (\ref{PolNitCorr:eqn}) were estimated  from the results for the two lowest $p_m$ bins.
Since the asymmetries on bound nucleons depend on kinematics (due to interference between different partial waves), we calculate a generous upper bound on the systematic uncertainty from this correction by varying  $A_{p}$ to plus or minus the maximum values consistent with Eq.~(\ref{eq:elasym}).

\subsubsection{Radiative Corrections}\label{section:radcor}
In order to compare observables like the double spin asymmetry reported in this paper to theoretical predictions, we must correct our measured results for radiative effects to convert them to the Born (one photon exchange) ones.  Both internal and external radiative (higher order electromagnetic) processes lead to a shift in kinematic variables like $Q^2$, $q$, and the direction of the $\vec{q}$ vector, which affect the extracted values for $p_m$ and $\cos\theta_{nq}$, and hence the asymmetry, through its kinematic dependence on these quantities. However, radiative effects on asymmetries tend to be smaller than on cross sections because the loss of events due to the ``radiative tail'' affects numerator and denominator similarly.

We determine the magnitude of these radiative effects by comparing a Monte Carlo simulation of the measured asymmetries with all radiative effects included to the same simulation with radiative effects turned off. This Monte Carlo simulation was run for beam energies of 1.6, 2.5, 4.2, and 5.7 GeV. Beam energies of 1.7, 5.6, and 5.8 GeV were not modeled since they were combined with similar beam energy runs and the difference in radiative effects for slightly different beam energies is very small.  We generated events distributed according to a  PWIA model for both the asymmetries and cross sections.  The initial proton momentum and polarization was chosen according to  probabilities calculated from the Argonne deuteron wave function, Ref.~\cite{roccoav18}, and the electron kinematics transformed into the rest frame of the proton. For the Born results, the Rosenbluth cross section and the asymmetry from Eq.~(\ref{eq:elasym}) were calculated and transformed back into the lab system. For radiated results, we used the full description of radiative effects in elastic scattering by Mo and Tsai~\cite{mo1969radiative} for the internal part, and calculated the effect of external bremsstrahlung on both the electron kinematics and polarization. We then applied a parametrization of our fiducial and kinematic cuts to select events within our acceptance. These events were then binned in the same bins as the real data and the asymmetries were calculated. The code for our simulation had been originally developed for the E6 experiment~\cite{Klimenko06} and has been extensively tested and compared to other cross section models. We also checked that the results of our simulation without radiative corrections agree closely with the asymmetries calculated from the model by Van Orden and Jeschonnek~\cite{Orden:Observables} for their PWIA case, confirming that our description of the scattering process in the Born approximation is in agreement with theory. 

We studied the systematic behavior of the difference between radiated and Born asymmetries from our Monte Carlo simulation, and found that in all cases, it could be described by a term proportional to the asymmetry (likely due to the change in effective virtual photon polarization) and a roughly constant offset. Therefore, we could write the desired Born asymmetry as 
\begin{align}
A_{||}^{Born} = \tau A_{||}^{meas} - \kappa ,
\end{align}
where the constants $\tau$ and $\kappa$ were determined from linear fits to our simulation results within each $Q^2$ bin and for each beam energy, separately for backward {\it vs.} sideway and forward spectator momenta. $\tau$ ranged from 0.95 to 1.28, and $\kappa$ ranged from -0.03 to 0.03. Overall, these corrections were small compared to the statistical uncertainties on the measured asymmetries (between 0.01 and 0.03 absolute, corresponding to less than 10\% of the asymmetry for most bins), and we estimated their systematic uncertainty by taking the full difference between radiated and Born asymmetries from our simulation.

\subsubsection{Systematic Uncertainties}\label{section:syserr}

In the previous sections, we described all corrections and conversion factors entering into the determination of the final Born (unradiated) double spin asymmetries for each kinematic bin. We also discussed our estimates for the systematic uncertainties on each of these corrections and conversion factors. We calculated the resulting systematic uncertainty due to each of these ingredients by varying one of them at a time (e.g., applying or not applying a correction, or varying factors within their uncertainties) and taking the difference between the extracted asymmetry due to this variation and the ``standard'' asymmetry for the nominal values and corrections. These differences were added in quadrature to determine the overall systematic uncertainty of each data point.

The contributions of these systematic uncertainties are shown in the plots in the following section as the outer error bars (systematic and statistical uncertainties added in quadrature). They typically range from about 40\% to 100\% of the statistical uncertainties, with a few outliers where both types of uncertainties are very large. The dominant contributions to the systematic uncertainties come from dilution factors (especially in the higher $p_m$ bins, where only a small fraction of the counts come from deuterium), corrections for polarized and unpolarized background contributions (again, most prominent at higher $p_m$), beam and target polarization (especially at the highest beam energy), and radiative corrections, in this order. We note that most of
these uncertainties (except for radiative corrections) depend on auxiliary measurements and therefore depend similarly on the total amount of collected data as statistical uncertainties. Furthermore, most corrections can vary significantly from one kinematic bin to the next, making the systematic uncertainties largely uncorrelated.

\section{RESULTS}\label{section:results}
After applying all corrections, the final physics (Born) asymmetry for $p_{m}$ bins 0 and 1 is
\begin{align}
&A_{||}(p_{m},Q^{2},\cos\theta_{nq})  =   \nonumber \\
& \frac{\tau}{ P_{b}P_{t}}\frac{(n^{+}-n^{-})-f_{back}A_{||}^{back}(n^{+}+n^{-})}{(1-f_{back})\left( a(n^{+}-n^{-})-b n_{B}\right)}-\kappa,
\label{fullasymintro1:eq}
\end{align}
where $n_{B}$ is the unpolarized background, $P_{b} P_{t}$ is the product of the beam and target polarizations, $\tau$ and $\kappa$ are correction terms associated with radiative corrections, and $a$ and $b$ are corrections terms for polarized background. The  Born asymmetry for $p_{m}$ bins 2, 3 and 4 is calculated as
\begin{align} 
 & A_{||}(p_{m},Q^{2},\cos\theta_{nq}) = \nonumber \\
& \frac{\tau}{ P_{b}P_{t}}\frac{(n^{+}-n^{-})-f_{back}A_{||}^{back}(n^{+}+n^{-})}{(1-f_{back})\left( (n^{+}-n^{-})-n_{B}\right)}-\kappa-C_{2}A_{p}  .
\label{fullasymintro2:eq}
\end{align}
The individual correction terms and their systematic uncertainties are explained in the previous section.  The resulting physics asymmetries and their statistical and systematic uncertainties were calculated for every data set and for every kinematic bin containing valid data.

\subsection{Combination of Asymmetries}\label{sec:combination}
For the final results, we combined the physics asymmetries for a given kinematic bin from different data sets with similar beam energies.  In all cases, we ascertained, using a student $t$-test, that the difference between the asymmetries in these data sets is small and consistent with statistical expectations. The asymmetries from different data sets were then averaged pairwise, using their (inverse squared) statistical uncertainties as weight.

First, data sets with different (opposite sign) target polarization but the same beam energy and the same torus polarity were combined.  Then, we combined asymmetries with similar energies and equal torus polarity.  For example, data set 1.6+ and 1.7+ were combined to form the 1.x+ data set and data set 1.6- and 1.7- were combined to form the 1.x- data set. This was also done for the 5.6 GeV, 5.7, and 5.8~GeV data sets  to form the 5.x GeV data set.  For our final combined values, we combined data sets with opposite torus polarity to obtain the four final data sets: 1.x, 2.5, 4.2, and 5.x.  

For our comparison with the theoretical models from Ref.~\cite{Orden:Observables}, we first calculated the predictions over a 
much finer grid in kinematic variables, including four values for the azimuth $\phi$ of the hadronic plane. The results were then averaged over each kinematic bin, using once again the statistical weight of the data from all data sets that contribute to a given bin. Hence, the data can be directly compared to these averaged predictions, with the same relative importance of all contributing kinematic points within a bin.

\subsection{Final Asymmetries}

\begin{figure}[!htb]
  \centering
\includegraphics[width=0.5\textwidth]{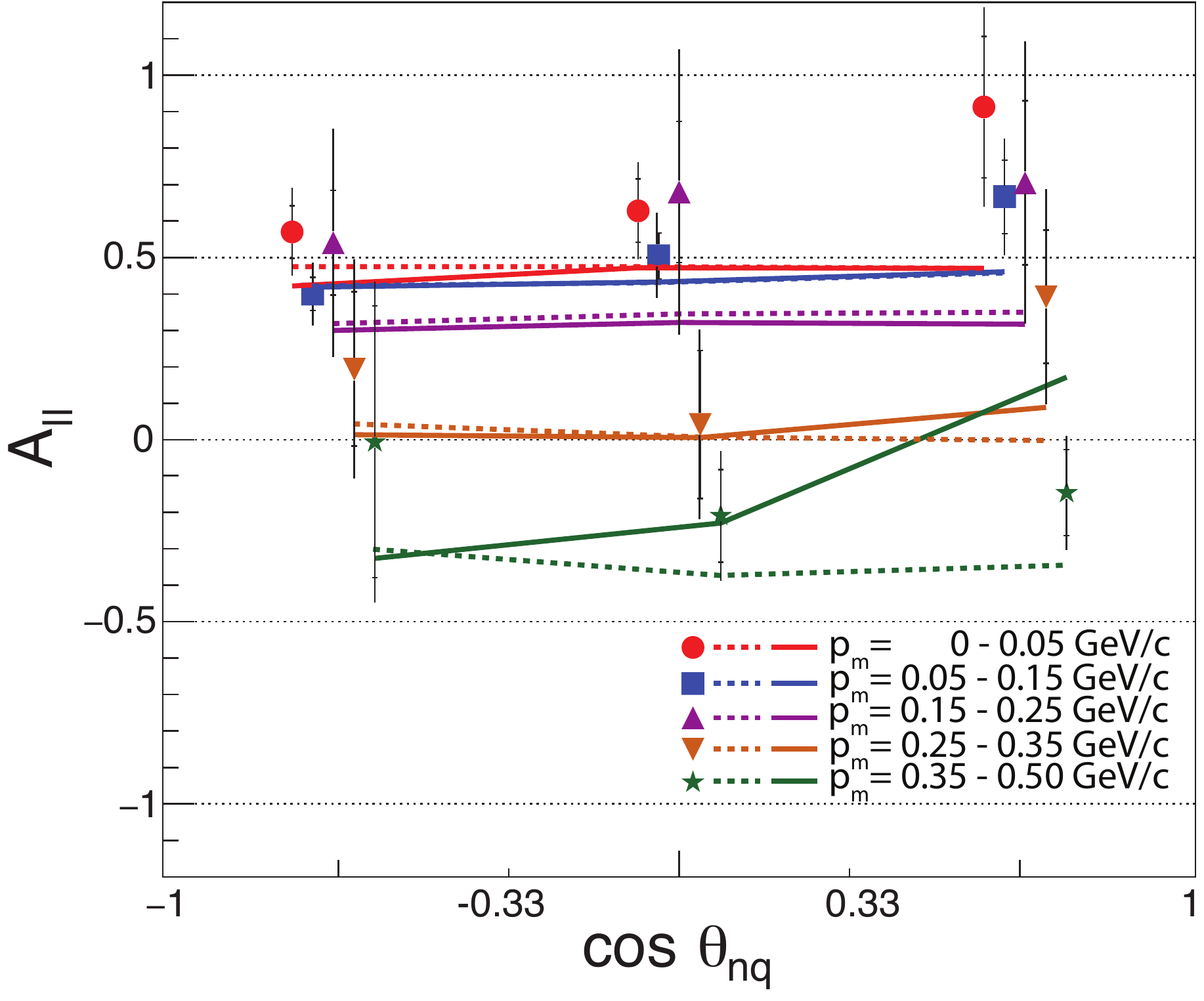}
  \caption{$A_{||}$ for beam energies of 1.6 -- 1.7 GeV and $0.38$ GeV$^2/c^2 \le Q^{2} \le 0.77$  GeV$^2/c^2$, vs. the cosine of the angle $\theta_{nq}$ between the direction of the virtual photon and the spectator neutron in the reaction \deep. The different symbols refer to different bins in missing momentum: red circles are for $p_m \le 0.05$ GeV/$c$, blue squares for $0.05$ GeV/$c$ $\le p_m \le 0.15$ GeV/$c$, purple triangles for $0.15$ GeV/$c$ $\le p_m \le 0.25$ GeV/$c$, orange inverted triangles for $0.25$ GeV/$c$ $\le p_m \le 0.35$ GeV/$c$, and green star symbols for $0.35$ GeV/$c$ $\le p_m \le 0.5$ GeV/$c$.  The inner error bars with horizontal risers indicate the statistical uncertainties, while the full error bars correspond to the statistical and systematic uncertainties added in quadrature. The dashed lines correspond to a PWIA prediction and the solid lines to a prediction including FSI, as explained in the text. They are color-coded for the same missing momentum bins as the data.}
  \label{16171:fig1}
\end{figure}

\begin{figure}[!htb]
  \centering
\includegraphics[width=0.5\textwidth]{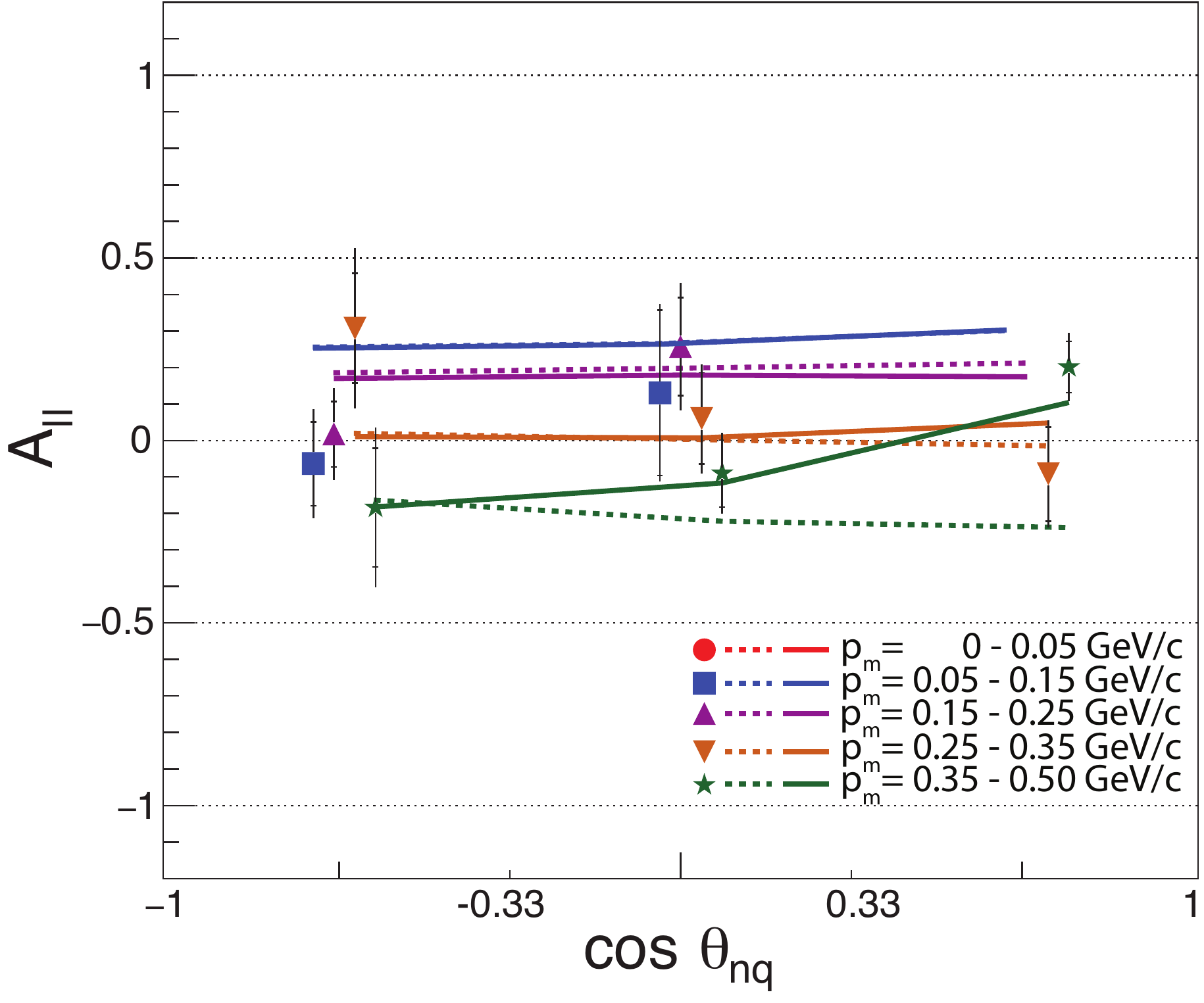}
  \caption{$A_{||}$ for a beam energy of 2.5 GeV and  the same $Q^2$ bin as before. All symbols and colors have the same meaning as in Fig.~\ref{16171:fig1}.}
  \label{251:fig1}
\end{figure}

Our final results for the double-spin asymmetry $A_{||}$ vs. $\cos \theta_{nq}$ and several missing momentum bins are presented in Figs.~\ref{16171:fig1} and \ref{251:fig1} for two specific beam energies and the same $Q^2$ bin. Tables of the complete results for all bins and beam energies can be found in the Appendix. Only results for bins fulfilling the criteria laid out in Sec.~\ref{binning:sec} are shown.

Our data show several of the expected features for $A_{||}$: at low missing momentum, the asymmetries are large and positive, and largely independent of $\cos \theta_{nq}$ within uncertainties. This is the kinematic domain where the struck proton is nearly on its energy shell, with asymmetries close to that for the free proton. Indeed, PWIA calculations (dashed lines in both Figures; see below) agree with this expectation and the data. As the missing momentum increases, the asymmetries deviate more strongly from the free proton ones, getting close to zero for $0.25$ GeV/$c$ $\le p_m \le 0.35$ GeV/$c$ and becoming even negative for our highest $p_m$ bin. From a na\"ive PWIA picture, this is to be expected, as higher proton (and therefore missing neutron) momenta correspond to the region where $S$- and $D$-state components of the deuteron wave function interfere or the $D$-State becomes even dominant. From simple Clebsch-Gordan arguments, it can be shown  that the average proton polarization inside deuterium is {\em negative} for this case, and becomes $- 1/2$ of the overall deuteron polarization for the $D$-state alone (it is $+ 1$ for a pure S-state). Again, this picture is supported by the PWIA calculations. However, some deviation from these expectations is seen in the $\cos \theta_{nq}$ -- dependence, which shows a tendency for the data points in $\cos \theta_{nq}$ bin 2 (with forward spectator momentum) to rise above the PWIA curves for the highest
$p_m$.  This effect is likely a consequence of FSI, as we explain in the following.

We compare our data to the Jeschonnek and Van Orden model~\cite{Orden:Observables}
for both the FSI and the PWIA case.  Two representative samples of the results can be seen in Figs.~\ref{16171:fig1} and \ref{251:fig1}.  The dashed lines indicate the result for PWIA only, while the solid lines correspond to the full calculation including FSI (see Sec.~\ref{section:theory}).  Each line has a color matching the color of the data in the corresponding $p_m$ bin.  It should be noted that for the 5.x GeV results, there was no model for FSI available yet and the results can only be compared with the PWIA model.  In Fig. \ref{16171:fig1}, it can be seen that there is very little difference between the FSI and PWIA model for the first three $p_m$ bins.  For the two highest $p_m$ bins, the two models predict different values as a function of $\cos\theta_{nq}$.  The FSI model predicts a more positive asymmetry in the forward $\cos\theta_{nq}$ bin than the PWIA model~\footnote{Large forward neutron momentum increases the likelihood that the neutron interacted with the struck proton, thereby increasing its momentum. Therefore, the asymmetry for these kinematics is more similar to that for lower missing momenta.}.  The same observations can be made in Fig.~\ref{251:fig1}.  The data show a similar trend, especially for $p_m$ bin 3 in Fig.~\ref{16171:fig1} and $p_m$ bin 4 in Fig.~\ref{251:fig1}, albeit somewhat less strongly (perhaps due to statistical fluctuations).

\begin{table}[!h] 
\centering
\caption[$\chi^{2}$ Values for PWIA and FSI Models]
{$\chi^{2}$ per degree of freedom of our data compared to a model~\cite{Orden:Observables} without (``PWIA'') and with (``FSI'') inclusion
of final state interaction effects.  All $\chi^{2}/dof$'s were calculated using all data points in a given Q$^2$ bin.  The 5.x data set could only be compared with the model using PWIA.  The 2.5 GeV and 4.2 GeV  data sets have few counts and therefore large (non-Gaussian) statistical uncertainty in the highest Q$^2$ bin 3, resulting in the low $\chi^{2}/dof$ values stated. $\chi^{2}$ values were calculated with the statistical and systematic uncertainty added in quadrature.}
\begin{tabular}{ccccc}
\hline
\hline
Energy$_{Beam}$ & $Q^2$ Bin & FSI $\chi^{2}/dof$ &  PWIA $\chi^{2}/dof$  & $dof$\\
\hline
1.x	&	0	&	2.406	&	2.576	&	9	\\
1.x	&	1	&	1.487	&	1.313	&	15	\\
1.x	&	2	&	1.409	&	1.981	&	7	\\
2.5	&	0	&	1.054	&	1.71	&	8	\\
2.5	&	1	&	1.523	&	4.817	&	10	\\
2.5	&	2	&	1.166	&	1.562	&	14	\\
2.5	&	3	&	0.584	&	0.543	&	7	\\
4.2	&	1	&	1.206	&	1.151	&	7	\\
4.2	&	2	&	1.097	&	1.212	&	8	\\
4.2	&	3	&	1.023	&	0.544	&	6	\\
5.x	&	2	&	n/a	&	0.456	&	5	\\
5.x	&	3	&	n/a	&	2.108	&	7	\\
\hline
\hline
\end{tabular}
\label{chi2sys:table}
\end{table}

We tested quantitatively  whether inclusion of FSI in the model improves the overall description of our data through a $\chi^2$ test for goodness of fit.  The $\chi^{2}/dof$ values were calculated for each $Q^2$ bin and beam energy as
\begin{equation}
 \chi^2/dof =\frac{\sum_{p_m,\cos\theta_{nq}}\frac{(A_{||}^{measured}-A_{||}^{theory})^2}{\sigma_{data}^2}}{N} , 
\label{funct:chi}
\end{equation}
where $N$ (degress of freedom, dof) is the number of data points summed over. Since most of our systematic uncertainties (due to polarized and unpolarized backgrounds, dilution factor, and radiative corrections) are largely uncorrelated bin to bin, we used the statistical and systematic uncertainties added in quadrature for the denominator, $\sigma_{data}$.  The values for $\chi^{2}/dof$ can be found in Table \ref{chi2sys:table}.  This table shows that the FSI model yields a lower $\chi^{2}/dof$ for most kinematic bins than  the PWIA model, sometimes drastically so.  The few bins with opposite trend either have very low $\chi^{2}/dof < 1$ for both models, or the difference is minimal.  The total value for $\chi^{2}$, summed over all bins excluding the 5.x GeV data, is 165.6 for the PWIA model ($dof = 91$, 
$p < 3 \cdot 10^{-6}$; or $\chi^{2} = 182.3$ for $dof = 103$ when we include the 5.x GeV bins) and $\chi^2 = 121$ ($dof = 91$,
$p \approx 0.02$) for the model with FSI included.  This difference %of nearly 45 units 
in $\chi^2$ indicates that the FSI model provides a significantly better description of the asymmetries overall than the PWIA model, in particular at high $p_m > 0.2$ GeV$/c$ where FSI effects are the largest and most of this difference arises. Conversely, at low $p_m$ the two models differ only by a little, making the PWIA description alone already a reasonably good one (within a few percent). For higher precision, or to cover higher $p_m$ bins, FSI must be included. The FSI model by Jeschonnek and Van Orden~\cite{Orden:Observables} appears to give a good description of the data over all kinematics, although the
agreement is not perfect.

\section{SUMMARY}\label{section:summary}
In summary, we have measured the exclusive double spin asymmetry $A_{||}$ for longitudinally polarized electrons scattering quasi-elastically on a deuteron target polarized along the beam direction, with simultaneous knocked-out proton detection, for 103 kinematic bins. Our data agree quite well with expectations of PWIA models for most bins, especially at lower missing momenta, $p_m < 0.2$ GeV/$c$.  Within errors, they also agree with previous measurements where they overlap in kinematics (see Sec.~\ref{section:wdata}; note that the variable $A^V_{ed}$ plotted in Figs.~\ref{aved:fig} and \ref{fig:BLAST} has the opposite sign convention as $A_{||}$). In particular, we see the decrease and even change in sign at higher missing momenta due to the increasing importance of the $D$-state component of the deuteron wave function. While our data are less precise and more sparse in missing momentum than those collected at NIKHEF~\cite{passchier} and BATES~\cite{degrush2010}, we cover a much larger range in $Q^2$ and beam energy as well as spectator momentum angle  $\cos \theta_{nq}$. We clearly see the effects of FSI in the dependence on this angle in several of our kinematic bins. Overall, our data are well described by a detailed theoretical model of the asymmetry~\cite{Orden:Observables} only if these FSI effects are properly included. They can serve to test future calculations as well as to provide better constraints for the extraction of neutron form factors and deuteron polarization (as a form of polarimetry for other processes) from quasi-elastic electron scattering.

\section*{Acknowledgments}
We would like to acknowledge the outstanding efforts of the staff of the Accelerator and the Physics Divisions at Jefferson Lab that made this experiment possible.  This material is based upon work supported by the U.S. Department of Energy, Office of Science, Office of Nuclear Physics under contracts DE-AC05-06OR23177, DE-FG02-96ER40960 and other contracts.  Jefferson Science Associates (JSA) operates the Thomas Jefferson National Accelerator Facility for the United States Department of Energy.  This work was supported in part by the U.S.~National Science Foundation, the Italian Instituto Nazionale di Fisica Nucleare,  the French Centre National de la Recherche Scientifique, the French Commissariat \`{a} l'Energie Atomique,  the Emmy Noether grant from the Deutsche Forschungs Gemeinschaft, the Scottish Universities Physics Alliance (SUPA), the United Kingdom's Science and Technology Facilities Council, the Chilean Comisi\'on Nacional de Investigaci\'on Cient\'ifica y Tecnol\'ogica (CONICYT), and the National Research Foundation of Korea.
 
\section*{Appendix}
This appendix contains tabular results for all measured kinematic bins.

\begin{table*}[!htb]																	
\caption{Measured asymmetries and bin-averages of the kinematic variables for beam energy 1.X GeV.}																	
\label{table:1xvalues}																	
\centering																	
\begin{tabular}{ccccccccc}																	
\hline																	
\hline																	
$Q^{2}$ Bin & $p_{m}$ Bin & cos $\theta_{nq}$ Bin & A$_{||}$ & $\sigma_{stat}$ & $\sigma_{sys}$ & $\bar{p}_{miss}$ & $\bar{Q}^2$ & cos $\bar{\theta}_{nq}$ \\																	
\hline																	
0	&	2	&	0	&	0.525	&	0.147	&	0.195	&	0.244	&	0.200	&	-0.670	\\
0	&	2	&	1	&	0.562	&	0.219	&	0.220	&	0.242	&	0.201	&	0.016	\\
0	&	2	&	2	&	-0.056	&	0.181	&	0.024	&	0.255	&	0.206	&	0.628	\\
0	&	3	&	0	&	0.246	&	0.103	&	0.060	&	0.235	&	0.293	&	-0.670	\\
0	&	3	&	1	&	0.093	&	0.108	&	0.037	&	0.241	&	0.296	&	0.039	\\
0	&	3	&	2	&	0.383	&	0.132	&	0.062	&	0.264	&	0.298	&	0.646	\\
0	&	4	&	0	&	0.071	&	0.099	&	0.034	&	0.231	&	0.404	&	-0.624	\\
0	&	4	&	1	&	-0.164	&	0.077	&	0.044	&	0.236	&	0.416	&	0.069	\\
0	&	4	&	2	&	-0.120	&	0.127	&	0.028	&	0.258	&	0.417	&	0.622	\\
1	&	0	&	0	&	0.570	&	0.073	&	0.057	&	0.617	&	0.037	&	-0.687	\\
1	&	0	&	1	&	0.628	&	0.086	&	0.076	&	0.610	&	0.037	&	-0.018	\\
1	&	0	&	2	&	0.913	&	0.194	&	0.217	&	0.605	&	0.036	&	0.662	\\
1	&	1	&	0	&	0.399	&	0.046	&	0.041	&	0.550	&	0.110	&	-0.692	\\
1	&	1	&	1	&	0.504	&	0.063	&	0.060	&	0.547	&	0.111	&	-0.014	\\
1	&	1	&	2	&	0.666	&	0.101	&	0.162	&	0.551	&	0.111	&	0.649	\\
1	&	2	&	0	&	0.540	&	0.144	&	0.169	&	0.514	&	0.195	&	-0.696	\\
1	&	2	&	1	&	0.679	&	0.194	&	0.199	&	0.510	&	0.198	&	0.022	\\
1	&	2	&	2	&	0.705	&	0.225	&	0.164	&	0.513	&	0.205	&	0.656	\\
1	&	3	&	0	&	0.194	&	0.211	&	0.093	&	0.497	&	0.290	&	-0.665	\\
1	&	3	&	1	&	0.041	&	0.203	&	0.058	&	0.497	&	0.296	&	0.060	\\
1	&	3	&	2	&	0.392	&	0.183	&	0.114	&	0.505	&	0.301	&	0.693	\\
1	&	4	&	0	&	-0.006	&	0.372	&	0.068	&	0.480	&	0.400	&	-0.590	\\
1	&	4	&	1	&	-0.209	&	0.128	&	0.057	&	0.497	&	0.413	&	0.083	\\
1	&	4	&	2	&	-0.146	&	0.118	&	0.066	&	0.519	&	0.416	&	0.726	\\
2	&	0	&	0	&	0.583	&	0.052	&	0.049	&	0.858	&	0.034	&	-0.678	\\
2	&	0	&	1	&	0.553	&	0.052	&	0.038	&	0.865	&	0.035	&	0.003	\\
2	&	0	&	2	&	0.690	&	0.049	&	0.047	&	0.874	&	0.035	&	0.686	\\
2	&	1	&	0	&	0.494	&	0.058	&	0.039	&	0.842	&	0.097	&	-0.663	\\
2	&	1	&	1	&	0.667	&	0.055	&	0.043	&	0.858	&	0.099	&	0.007	\\
2	&	1	&	2	&	0.655	&	0.046	&	0.041	&	0.878	&	0.099	&	0.685	\\
2	&	4	&	2	&	0.601	&	0.425	&	0.128	&	0.844	&	0.425	&	0.765	\\
\hline																	
\hline																	
\end{tabular}																	
\end{table*}

\begin{table*}[!htb]																	
\caption{Measured asymmetries and bin-averages of the kinematic variables for beam energy 2.5 GeV.}																	
\label{table:25values}																	
\centering																	
\begin{tabular}{ccccccccc}																	
\hline																	
\hline																	
$Q^{2}$ Bin & $p_{m}$ Bin & cos $\theta_{nq}$ Bin & A$_{||}$ & $\sigma_{stat}$ & $\sigma_{sys}$ & $\bar{p}_{miss}$ & $\bar{Q}^2$ & cos $\bar{\theta}_{nq}$ \\																	
\hline																	
0	&	2	&	0	&	-0.103	&	0.134	&	0.051	&	0.234	&	0.202	&	-0.666	\\
0	&	2	&	1	&	0.175	&	0.169	&	0.067	&	0.232	&	0.203	&	0.017	\\
0	&	3	&	0	&	0.062	&	0.083	&	0.025	&	0.232	&	0.295	&	-0.675	\\
0	&	3	&	1	&	0.143	&	0.098	&	0.057	&	0.236	&	0.297	&	0.034	\\
0	&	3	&	2	&	-0.088	&	0.132	&	0.012	&	0.262	&	0.298	&	0.644	\\
0	&	4	&	0	&	-0.185	&	0.099	&	0.137	&	0.230	&	0.405	&	-0.636	\\
0	&	4	&	1	&	0.008	&	0.081	&	0.031	&	0.237	&	0.415	&	0.056	\\
0	&	4	&	2	&	0.259	&	0.124	&	0.084	&	0.261	&	0.415	&	0.620	\\
1	&	1	&	0	&	-0.064	&	0.115	&	0.062	&	0.559	&	0.115	&	-0.683	\\
1	&	1	&	1	&	0.131	&	0.226	&	0.019	&	0.544	&	0.115	&	-0.010	\\
1	&	2	&	0	&	0.017	&	0.089	&	0.038	&	0.540	&	0.199	&	-0.698	\\
1	&	2	&	1	&	0.257	&	0.135	&	0.041	&	0.528	&	0.201	&	0.018	\\
1	&	3	&	0	&	0.308	&	0.150	&	0.070	&	0.527	&	0.293	&	-0.682	\\
1	&	3	&	1	&	0.060	&	0.125	&	0.025	&	0.523	&	0.298	&	0.049	\\
1	&	3	&	2	&	-0.092	&	0.129	&	0.020	&	0.525	&	0.302	&	0.688	\\
1	&	4	&	0	&	-0.184	&	0.163	&	0.070	&	0.522	&	0.404	&	-0.625	\\
1	&	4	&	1	&	-0.090	&	0.092	&	0.026	&	0.525	&	0.415	&	0.077	\\
1	&	4	&	2	&	0.201	&	0.070	&	0.029	&	0.537	&	0.418	&	0.716	\\
2	&	0	&	0	&	0.553	&	0.055	&	0.060	&	1.176	&	0.036	&	-0.684	\\
2	&	0	&	1	&	0.582	&	0.055	&	0.043	&	1.192	&	0.036	&	-0.007	\\
2	&	0	&	2	&	0.482	&	0.055	&	0.035	&	1.219	&	0.035	&	0.674	\\
2	&	1	&	0	&	0.346	&	0.044	&	0.031	&	1.064	&	0.106	&	-0.689	\\
2	&	1	&	1	&	0.502	&	0.049	&	0.053	&	1.098	&	0.106	&	-0.012	\\
2	&	1	&	2	&	0.454	&	0.050	&	0.073	&	1.142	&	0.105	&	0.661	\\
2	&	2	&	0	&	0.495	&	0.160	&	0.162	&	1.015	&	0.194	&	-0.684	\\
2	&	2	&	1	&	0.418	&	0.224	&	0.101	&	1.037	&	0.196	&	0.008	\\
2	&	2	&	2	&	0.464	&	0.269	&	0.090	&	1.044	&	0.202	&	0.663	\\
2	&	3	&	0	&	0.305	&	0.272	&	0.069	&	0.989	&	0.290	&	-0.658	\\
2	&	3	&	1	&	0.315	&	0.302	&	0.079	&	1.003	&	0.297	&	0.042	\\
2	&	4	&	0	&	-0.483	&	0.395	&	0.107	&	0.967	&	0.401	&	-0.604	\\
2	&	4	&	1	&	-0.534	&	0.267	&	0.113	&	0.994	&	0.415	&	0.080	\\
2	&	4	&	2	&	0.531	&	0.218	&	0.134	&	1.004	&	0.422	&	0.734	\\
3	&	0	&	0	&	0.708	&	0.127	&	0.082	&	1.729	&	0.035	&	-0.675	\\
3	&	0	&	1	&	0.750	&	0.113	&	0.049	&	1.731	&	0.035	&	0.000	\\
3	&	0	&	2	&	0.699	&	0.110	&	0.046	&	1.737	&	0.035	&	0.685	\\
3	&	1	&	0	&	0.661	&	0.140	&	0.072	&	1.709	&	0.098	&	-0.666	\\
3	&	1	&	1	&	0.601	&	0.113	&	0.034	&	1.727	&	0.101	&	0.022	\\
3	&	1	&	2	&	0.532	&	0.091	&	0.035	&	1.746	&	0.103	&	0.695	\\
3	&	2	&	0	&	-0.119	&	0.499	&	0.059	&	1.685	&	0.189	&	-0.639	\\
\hline																	
\hline																	
\end{tabular}																	
\end{table*}

\begin{table*}[!htb]																	
\caption{Measured asymmetries and bin-averages of the kinematic variables for beam energy 4.2 GeV.}																	
\label{table:42values}																	
\centering																	
\begin{tabular}{ccccccccc}																	
\hline																	
\hline																	
$Q^{2}$ Bin & $p_{m}$ Bin & cos $\theta_{nq}$ Bin & A$_{||}$ & $\sigma_{stat}$ & $\sigma_{sys}$ & $\bar{p}_{miss}$ & $\bar{Q}^2$ & cos $\bar{\theta}_{nq}$ \\																	
\hline																	
1	&	2	&	0	&	0.313	&	0.248	&	0.071	&	0.545	&	0.202	&	-0.693	\\
1	&	3	&	0	&	-0.066	&	0.336	&	0.081	&	0.536	&	0.294	&	-0.693	\\
1	&	3	&	1	&	-0.022	&	0.271	&	0.034	&	0.532	&	0.298	&	0.045	\\
1	&	3	&	2	&	-0.196	&	0.314	&	0.064	&	0.533	&	0.302	&	0.687	\\
1	&	4	&	0	&	0.545	&	0.315	&	0.224	&	0.530	&	0.405	&	-0.648	\\
1	&	4	&	1	&	0.368	&	0.204	&	0.115	&	0.530	&	0.415	&	0.073	\\
1	&	4	&	2	&	-0.169	&	0.211	&	0.047	&	0.547	&	0.417	&	0.713	\\
2	&	0	&	0	&	0.422	&	0.145	&	0.076	&	1.243	&	0.037	&	-0.686	\\
2	&	0	&	1	&	0.501	&	0.161	&	0.094	&	1.229	&	0.037	&	-0.016	\\
2	&	1	&	0	&	0.299	&	0.093	&	0.043	&	1.128	&	0.111	&	-0.698	\\
2	&	1	&	1	&	0.538	&	0.134	&	0.156	&	1.126	&	0.111	&	-0.019	\\
2	&	2	&	0	&	0.710	&	0.228	&	0.121	&	1.055	&	0.197	&	-0.705	\\
2	&	2	&	1	&	-0.112	&	0.323	&	0.151	&	1.053	&	0.199	&	0.004	\\
2	&	4	&	1	&	0.070	&	0.297	&	0.063	&	1.031	&	0.415	&	0.074	\\
2	&	4	&	2	&	-0.032	&	0.378	&	0.167	&	1.030	&	0.422	&	0.726	\\
3	&	0	&	0	&	0.436	&	0.087	&	0.048	&	2.025	&	0.035	&	-0.682	\\
3	&	0	&	1	&	0.514	&	0.087	&	0.056	&	2.020	&	0.035	&	0.000	\\
3	&	0	&	2	&	0.477	&	0.085	&	0.050	&	2.034	&	0.035	&	0.680	\\
3	&	1	&	0	&	0.539	&	0.087	&	0.056	&	2.025	&	0.102	&	-0.682	\\
3	&	1	&	1	&	0.469	&	0.079	&	0.053	&	2.032	&	0.102	&	0.000	\\
3	&	1	&	2	&	0.383	&	0.076	&	0.049	&	2.074	&	0.102	&	0.674	\\
\hline																	
\hline																	
\end{tabular}																	
\end{table*} 																	
																	
\begin{table*}[!htb]																	
\caption{Measured asymmetries and bin-averages of the kinematic variables for beam energy 5.x GeV.}																	
\label{table:5xvalues}																	
\centering																	
\begin{tabular}{ccccccccc}																	
\hline																	
\hline																	
$Q^{2}$ Bin & $p_{m}$ Bin & cos $\theta_{nq}$ Bin & A$_{||}$ & $\sigma_{stat}$ & $\sigma_{sys}$ & $\bar{p}_{miss}$ & $\bar{Q}^2$ & cos $\bar{\theta}_{nq}$ \\																	
\hline																	
2	&	1	&	0	&	0.152	&	0.056	&	0.206	&	1.190	&	0.113	&	-0.699	\\
2	&	1	&	1	&	0.298	&	0.385	&	0.217	&	1.170	&	0.112	&	-0.019	\\
2	&	2	&	0	&	0.487	&	0.098	&	0.158	&	1.118	&	0.198	&	-0.708	\\
%2	&	2	&	2	&	-10.976	&	0.066	&	0.181	&	1.102	&	0.205	&	0.667	\\
2	&	4	&	1	&	0.432	&	0.308	&	-0.231	&	1.081	&	0.416	&	0.071	\\
3	&	0	&	0	&	0.088	&	0.059	&	0.296	&	2.079	&	0.036	&	-0.682	\\
3	&	0	&	1	&	0.088	&	0.061	&	0.297	&	2.112	&	0.036	&	-0.003	\\
3	&	0	&	2	&	0.088	&	0.062	&	0.296	&	2.136	&	0.036	&	0.676	\\
3	&	1	&	0	&	0.077	&	0.036	&	0.301	&	2.037	&	0.103	&	-0.683	\\
3	&	1	&	1	&	0.076	&	0.040	&	0.303	&	2.078	&	0.103	&	-0.005	\\
3	&	1	&	2	&	0.078	&	0.036	&	0.293	&	2.149	&	0.103	&	0.672	\\
3	&	2	&	0	&	0.358	&	0.124	&	0.249	&	2.015	&	0.194	&	-0.681	\\
\hline																	
\hline																	
\end{tabular}																	
\end{table*} 																	
%\bibliography{QEDSA}
%merlin.mbs apsrev4-1.bst 2010-07-25 4.21a (PWD, AO, DPC) hacked
%Control: key (0)
%Control: author (8) initials jnrlst
%Control: editor formatted (1) identically to author
%Control: production of article title (-1) disabled
%Control: page (0) single
%Control: year (1) truncated
%Control: production of eprint (0) enabled
%

\end{document}